\documentclass[sigconf, 10pt, screen]{acmart}
\pdfoutput=1

\renewcommand\footnotetextcopyrightpermission[1]{} 

\newif\ifdebugdoc\debugdocfalse
\newif\ifflow\flowfalse

\makeatletter
\def\subsubsection{\@startsection{subsubsection}{3}{10pt}%
                                 {-.5\baselineskip \@plus -2\p@ \@minus -.2\p@}%
                 {2.5\p@}{\@subsubsecfont}}
\makeatother

\ifdebugdoc
    \definecolor{ForestGreen}{rgb}{0.0, 0.27, 0.13}
    \definecolor{DarkPurple}{rgb}{0.34, 0.03, 0.38}
    \newcommand{\outline}[1]{\textbf{\colorbox{yellow}{Outline:}\textcolor{red}{#1.}}}

    \newcommand{\add}[1]{\textcolor{red}{#1}}
    \newcommand{\del}[1]{\textcolor{blue}{\sout{#1}}}
    \newcommand{\todo}[1]{\textcolor{red}{{\bf TODO}: {#1}}}
    \newcommand{\save}[1]{}
    \newcommand{\yuanjie}[1]{\textcolor{red}{{\bf Yuanjie}: {#1}}}
    \newcommand{\lixin}[1]{\textcolor{blue}{{\bf Lixin}: {#1}}}
     \newcommand{\wei}[1]{\textcolor{cyan}{{\bf Wei}: {#1}}}
      \newcommand{\zhaowei}[1]{\textcolor{blue}{{\bf Zhaowei}: {#1}}}
      \newcommand{\yimei}[1]{\textcolor{purple}{{\bf Yimei}: {#1}}}
      \newcommand{\ouyang}[1]{\textcolor{blue}{{\bf Ouyang}: {#1}}}
      \newcommand{\junyi}[1]{\textcolor{cyan}{{\bf Junyi}: {#1}}}
    \newcommand{\fixme}[1]{\textcolor{blue}{{\bf FIXME}: {#1}}}
    \newcommand{\question}[1]{\textcolor{purple}{{\bf Question}: {#1}}}
\else
    \newcommand{\outline}[1]{}

    \newcommand{\add}[1]{#1}
    \newcommand{\del}[1]{}
    \newcommand{\todo}[1]{}
    \newcommand{\save}[1]{}
    \newcommand{\yuanjie}[1]{}
    \newcommand{\lixin}[1]{}
    \newcommand{\wei}[1]{}
    \newcommand{\zhaowei}[1]{}
    \newcommand{\yimei}[1]{}
    \newcommand{\ouyang}[1]{}
    \newcommand{\junyi}[1]{}
    \newcommand{\fixme}[1]{}
	\newcommand{\question}[1]{}
\fi

\ifflow
    \newcommand{\p}[1]{\vskip 1ex\noindent\colorbox{yellow}{\parbox{\columnwidth}{\textbf{Point:} {#1}}}}
    \newcommand{\key}[1]{\vskip 1ex\noindent\colorbox{yellow}{\parbox{\columnwidth}{\textbf{Keys:} {#1}}}}
    \newcommand{\q}[1]{\vskip 1ex\noindent\colorbox{cyan}{\parbox{\columnwidth}{\textbf{Question:} {#1}}}}
\else
    \newcommand{\p}[1]{}
    \newcommand{\key}[1]{}
    \newcommand{\q}[1]{}
\fi

\newcommand{\paragraphb}[1]{\vspace{0mm}\noindent\textbf{#1}}

\def\ie{i.e.\xspace}
\def\eg{e.g.\xspace}

\usepackage[export]{adjustbox}
\usepackage{url}
\usepackage{subfig}
\usepackage{xcolor}
\usepackage{xspace}
\usepackage{amsmath}
\usepackage{balance}
\usepackage{booktabs}
\usepackage{tabularx}
\usepackage{multirow}
\usepackage{multirow}
\usepackage{graphicx}
\usepackage{amsfonts}
\usepackage{verbatim}
\usepackage{algorithm}
\usepackage{algpseudocode}
\makeatletter
\renewcommand{\ALG@beginalgorithmic}{\scriptsize}
\makeatother
\usepackage[normalem]{ulem}
\usepackage{paralist}
\usepackage{slashbox}
\usepackage{varwidth}
\usepackage{amsmath}
\usepackage[acronym]{glossaries}
\usepackage{tablefootnote}

\algnewcommand\algorithmicswitch{\textbf{switch}}
\algnewcommand\algorithmiccase{\textbf{case}}
\algnewcommand\algorithmicassert{\texttt{assert}}
\algnewcommand\Assert[1]{\State \algorithmicassert(#1)}%
\algdef{SE}[SWITCH]{Switch}{EndSwitch}[1]{\algorithmicswitch\ #1\ \algorithmicdo}{\algorithmicend\ \algorithmicswitch}%
\algdef{SE}[CASE]{Case}{EndCase}[1]{\algorithmiccase\ #1}{\algorithmicend\ \algorithmiccase}%
\algtext*{EndSwitch}%
\algtext*{EndCase}%

\algnewcommand{\IIf}[1]{\State\algorithmicif\ #1\ \algorithmicthen}
\algnewcommand{\EndIIf}{\unskip\ \algorithmicend\ \algorithmicif}

\theoremstyle{plain}
  
  \newtheorem{prop}{Proposition}
  
  \newtheorem{definition}{Definition}

\theoremstyle{acmdefinition}

\theoremstyle{remark}

\usepackage{mathtools}

\usepackage[skip=3pt]{caption}
\usepackage{enumerate}
\usepackage{enumitem}
\setlist[enumerate,1]{label=(\arabic*).,font=\textup,
leftmargin=7mm,labelsep=1.5mm,topsep=0mm,itemsep=-0.8mm}
\setlist[enumerate,2]{label=(\alph*).,font=\textup,
leftmargin=7mm,labelsep=1.5mm,topsep=-0.8mm,itemsep=-0.8mm}

\begin{document}
\title{
Instability of {Self-Driving} Satellite Mega-Constellation
}
\subtitle{From Theory to Practical Impacts on Network Lifetime and Capacity}

\author{\fontsize{13}{15}\selectfont Yimei Chen, Yuanjie Li, Hewu Li, Lixin Liu, Li Ouyang, Jiabo Yang, Junyi Li,\\ Jianping Wu, Qian Wu, Jun Liu, Zeqi Lai}

\affiliation{
	\institution{
	Institute for Network Sciences and Cyberspace, Tsinghua University, Beijing 100084, China}
	\city{}
	\country{}
}

\pagestyle{plain} 

\begin{abstract}


Low Earth Orbit (LEO) satellite mega-constellations aim to enable high-speed Internet for numerous users anywhere on Earth.
To safeguard their network infrastructure in congested outer space, they perform automatic orbital maneuvers to avoid collisions with {external} debris and satellites.
However, 
our control-theoretic analysis and empirical validation using Starlink's space situational awareness datasets discover that, 
these safety-oriented maneuvers themselves can threaten safety and networking via
{\em cascaded collision avoidance inside the mega-constellation}.
This domino effect forces a dilemma between long-term LEO network lifetime and short-term LEO network capacity.
Its root cause is that, 
the decades-old {\em local pairwise} maneuver paradigm for standalone satellites
is inherently unstable if scaled out to recent mega-constellation networks.
We thus propose an alternative bilateral maneuver control that 
stabilizes self-driving mega-constellations for concurrent network lifetime {\em and} capacity boosts.
Our operational trace-driven emulation shows a \add{8$\times$} network lifetime extension in Starlink without limiting its network capacity.

\end{abstract} 

\maketitle

\section{Introduction}
\label{sec:intro}

\begin{figure}[t]
	\centering
	\vspace{-2.5mm}
	\subfloat[Space congestion]{
		\hspace{-1mm}
		\includegraphics[width=0.305\columnwidth]{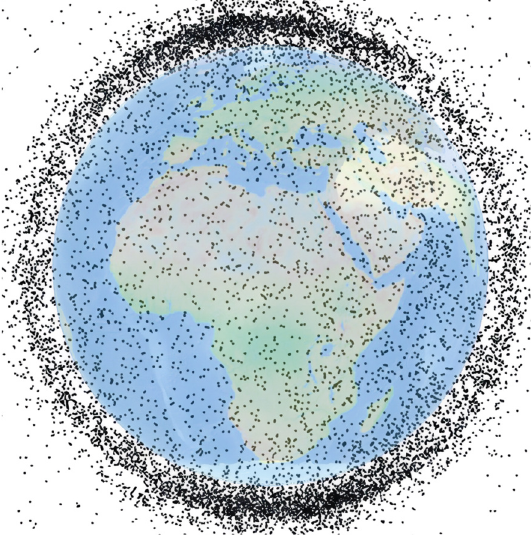}
		\label{fig:harsh-space}
	}
	\subfloat[Maneuvers' impacts on satellite networks]{
		\hspace{-3.2mm}
		\includegraphics[width=0.67\columnwidth]{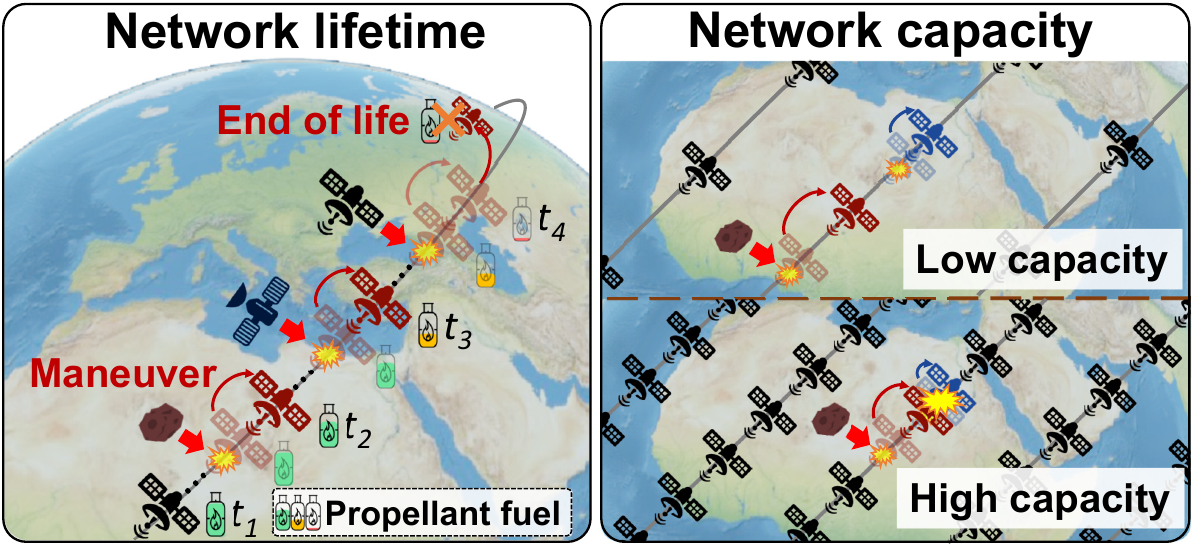}
		\label{fig:negative-impacts}
	}
	\vspace{-3mm}
	\subfloat[Starlink's constellation scale and maneuver growth over time]{
	\includegraphics[width=0.505\columnwidth]{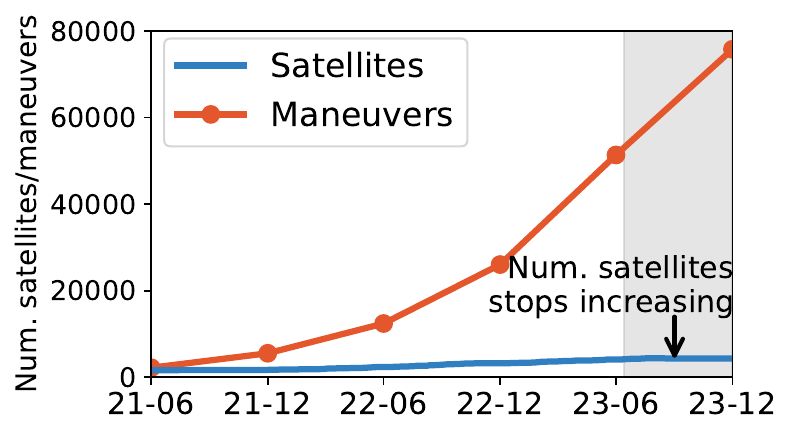}
	\includegraphics[width=0.495\columnwidth]{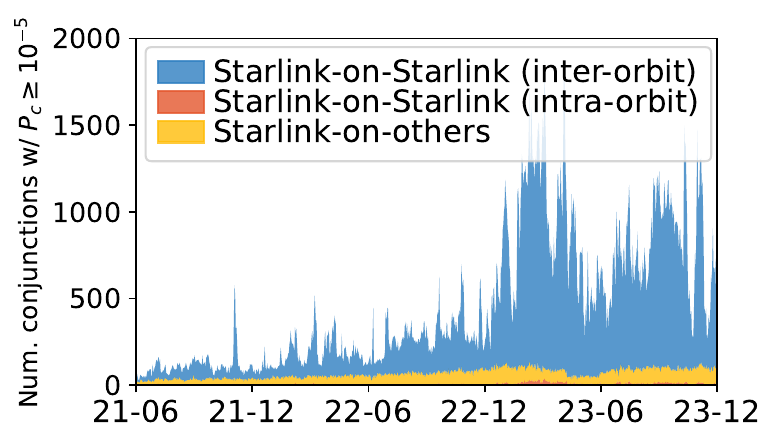}
	\label{fig:semi-report_a}
	\label{fig:conjunction-sos}
	}
	\caption{
		Unstable automatic orbital maneuvers and their impacts on satellite mega-constellation networks.
	}
	\label{fig:overview}
	\label{fig:semi-reports}
	\vspace{-2mm}
\end{figure}

Low Earth Orbit (LEO) satellite mega-constellation networks are revolutionizing the future Internet. 
Powered by 1,000--10,000s satellites with ultrahigh-capacity wireless links,
they aim to enable broadband Internet access 
to the remaining 2.6 billion ``unconnected'' users \cite{underserved-user} anywhere on Earth.
Thanks to recent technological advances in satellite miniaturization and rocket reusability,
this vision has gained traction from considerable
industrial efforts like Starlink \cite{starlink}, OneWeb \cite{oneweb}, Amazon Kuiper \cite{kuiper}, Iridium \cite{iridium-constellation}, Inmarsat \cite{inmarsat-leo},
and more.

In this work, we ask a simple question:
{\em Can LEO mega-constellations retain sustainable high-speed Internet access to numerous users at scale?}
Our study is motivated by the rule of thumb that, 
while a large system like the mega-constellation network can expand its capacity by adding more satellites,
such a scale-out expansion is often associated with
{\em inherent system complexity hurdles}
for its scalability and robustness.  
It is open to question whether mega-constellations 
can always sustain their rapid network capacity expansions today to 
serve more and more users for revenues and social goods.

This paper studies the question from a unique perspective of LEO mega-constellation networks: {\em autonomous driving}.
Compared to terrestrial networks, LEO satellites operate in the harsh, congested outer space.
As shown in Figure~\ref{fig:harsh-space}, they are surrounded by countless space debris and third-party satellites \cite{debris-number}.
This leads to unprecedented satellite collision risks for the LEO network infrastructure.
To this end, the latest LEO mega-constellations have installed self-driving systems for their satellites to sense space collision risks and avoid them via maneuvers \cite{starlink-maneuver, starlink-maneuver-2}.
So far, these systems have successfully safeguarded
mega-constellations via 
10,000s of maneuvers every half year
\cite{starlink-semi-annual-2021-7,starlink-semi-annual-2021-12,starlink-semi-annual-2022-7,starlink-semi-annual-2022-12,starlink-semi-annual-2023-6,starlink-semi-annual-2023-12} (Figure~\ref{fig:semi-report_a}), thus seeming to facilitate reliable and performant network infrastructure. 

Unfortunately, we find that these self-driving systems can be a double-edged sword for safety and networking.
While excellent in avoiding collisions with {\em external} debris/satellites,
they can also amplify {\em internal} satellite collision risks inside a mega-constellation. 
This phenomenon has been evidenced in Starlink's official semi-annual reports \cite{starlink-semi-annual-2021-7,starlink-semi-annual-2021-12,starlink-semi-annual-2022-7,starlink-semi-annual-2022-12,starlink-semi-annual-2023-6,starlink-semi-annual-2023-12} and our empirical results in Figure~\ref{fig:semi-report_a} (detailed in $\S$\ref{sec:motivation}). 
They unveil an explosive growth of Starlink's maneuvers that significantly surpasses its satellite growth.
This explosive maneuver growth can magnify satellite network connectivity/service interruptions \cite{mobicom23li,peiro2000galileo,esa-gsl-disruption-in-maneuver,starlink-degraded-service-congestion}
and distortion of nominal satellite distributions in the mega-constellation  \cite{fan2017ground,aorpimai2007repeat,kim2021maintaining}, both being
well-known threats to satellite missions.
Surprisingly, most maneuvers occur among Starlink's own satellites, 
contrasting with its claim in \cite{starlink-maneuver-2} that its careful mega-constellation design has deconflicted most internal satellite collisions. 

To understand this counter-intuitive phenomenon, 
we propose a control-theoretic framework to characterize the structural properties of automatic orbital maneuvers in mega-constellations and their impacts on networking.
We formally prove that this 
phenomenon is rooted in the {\em endogenous instability} of self-driving mega-constellations (rather than specific to Starlink only). 
To scale space collision avoidance to numerous satellites,
most operators follow the decades-old  {\em local pairwise} maneuver paradigm \cite{krage2020nasa, operator-handbook, alfano2005numerical,probe2022prototype,brown2014simulated}:
Each satellite will only maneuver in response to its most risky upcoming conjunction with another satellite/debris each time.
While reasonable and cost-effective for each satellite,
this local maneuver's overall global effect is the undesirable {\bf cascaded collision avoidance}:
A maneuver for external collision avoidance 
will move a satellite closer to its neighboring satellites inside the mega-constellation, which increases its internal collision risk and forces its neighboring satellite to maneuver further. 
This domino effect amplifies hop-by-hop inside the mega-constellation, leading to the explosive maneuver growth in Figure~\ref{fig:semi-report_a}.
Our control-theoretic analysis in $\S$\ref{sec:theory} and empirical validation using Starlink's large-scale space situational awareness datasets in $\S$\ref{sec:validation} show that, 
this instability threatens LEO networks in at least two aspects (Figure~\ref{fig:negative-impacts}):

\begin{compactenum}
\item {\bf Network lifetime:} 
Each satellite only has limited fuel to perform finite maneuvers (\eg, 350 times in Starlink \cite{starlink-maneuver-lifetime}), 
after which it is permanently deorbited for space safety \cite{starlink-maneuver,starlink-maneuver-2}.
Cascaded maneuvers exhaust satellites' budgets, shorten their life for network services \cite{starlink-satellite-lifetime}, and raise operators' satellite replacement costs; 

\item {\bf Network capacity:}
To save the LEO network lifetime, one could configure its local pairwise maneuver policy for stabilization. 
However, we prove this effort will unavoidably increase collision risks and force larger inter-satellite safe distance, thus throttling the satellite network scale and capacity to serve more users.
\end{compactenum}

\begin{figure}[t]
\centering
\vspace{-4mm}
\subfloat[
LEO satellite constellation
]{
\includegraphics[width=0.45\columnwidth]{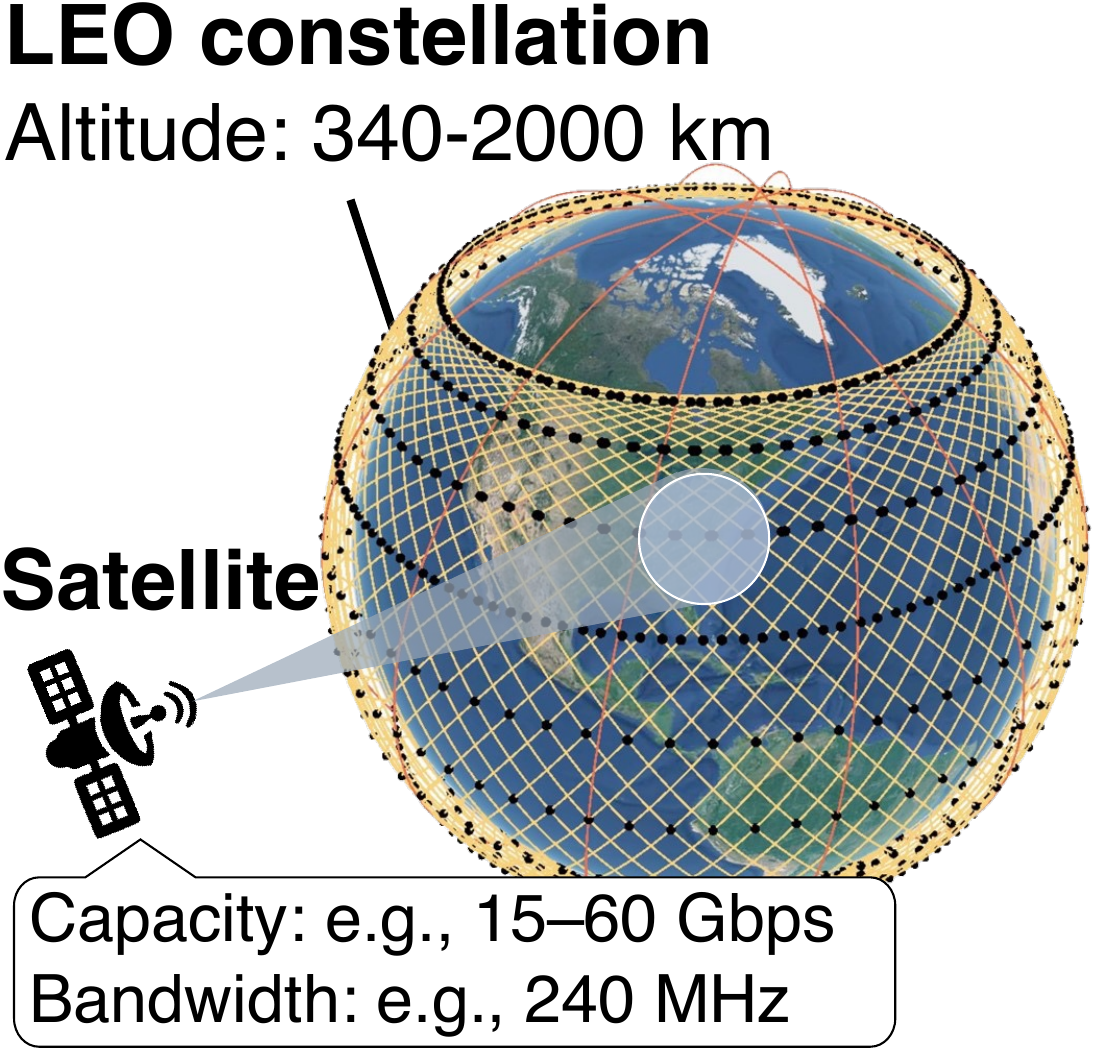}
\label{fig:background}
}
\subfloat[
The need for mega-constellation.
]{
\includegraphics[width=0.55\columnwidth]{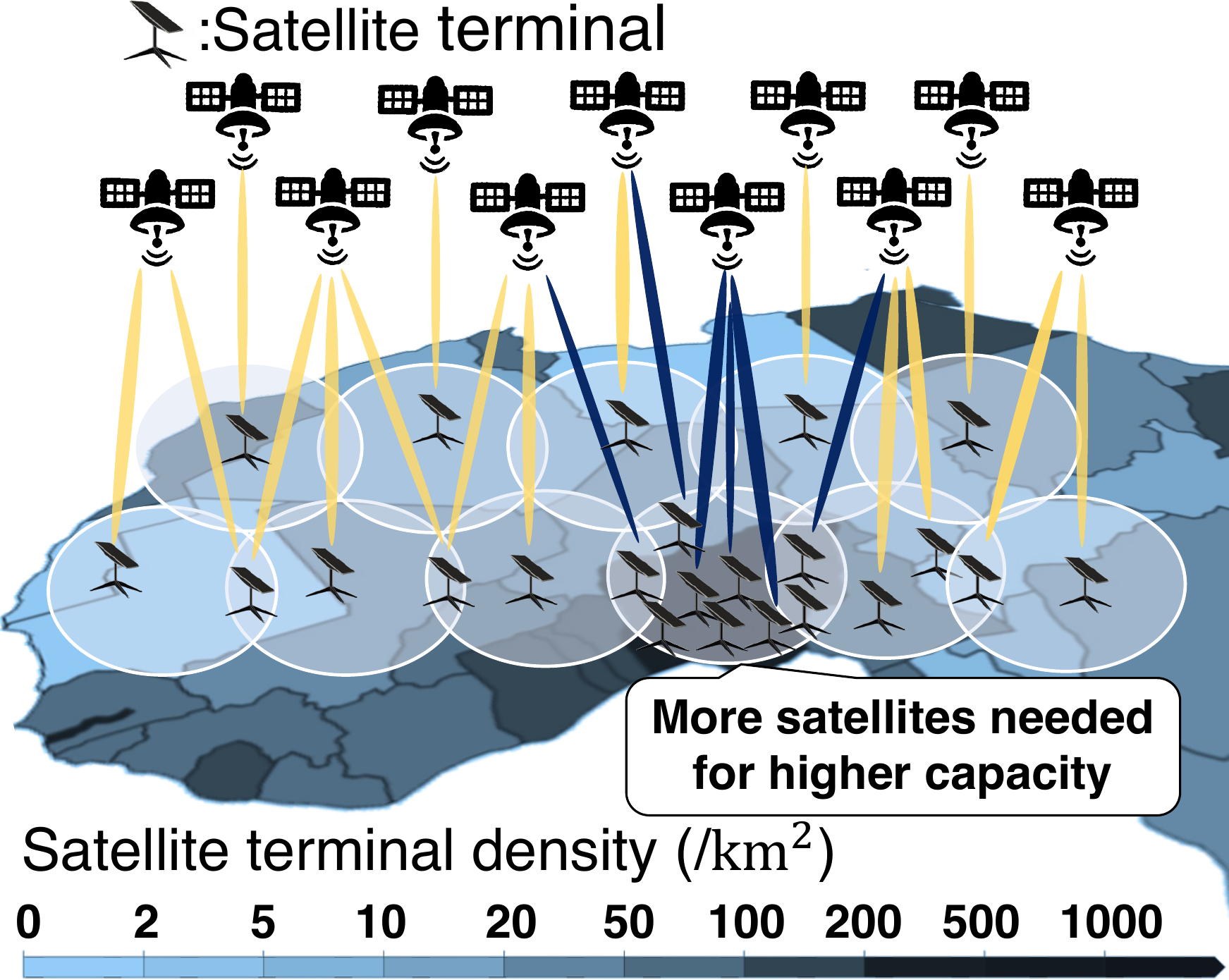}
}
\caption{LEO satellite mega-constellation network. 
}
\label{fig:back}
\label{fig:coverage}
\vspace{-2mm}
\end{figure}


Our study further sheds light on how to avoid unstable maneuvers and bypass the above dilemma between LEO network lifetime and capacity. 
The key is to depart from local pairwise maneuvers to global maneuver coordination among satellites. 
Its challenge, however, is the coordination complexity when scaled out to the mega-constellation. 
To this end, we simplify the global maneuver coordination to the distributed, network-friendly bilateral maneuver control ($\S$\ref{sec:solution}).  
By incrementally adding a local backward control to the existing pairwise maneuvers, 
this method can provably decouple local collision avoidance (which limits LEO network capacity) from global maneuver stability (which impacts LEO network lifetime) for their concurrent satisfaction.
Our operational trace-driven emulation shows that it can expand Starlink's lifetime by \add{8$\times$} without limiting its total capacity.

\paragraphb{Ethics: }This work does not raise any ethical issues.


\section{Self-Driving Mega-Constellation}
\label{sec:back}

We introduce the LEO mega-constellation networks ($\S$\ref{sec:back:leo-network}),
their congested operation environment in outer space ($\S$\ref{sec:back:congested-space}),
and how they enforce safety via automatic maneuvers ($\S$\ref{sec:back:collision-avoidance}).

\subsection{The LEO Mega-Constellation Network}
\label{sec:back:leo-network}


LEO networks complement terrestrial networks to enable broadband Internet access 
in under-served areas. 
As shown in Figure~\ref{fig:back}, 
each LEO satellite runs at an altitude of 340--2,000 km. 
It equips ultrahigh-capacity radio links 
(\eg, 240 MHz Ku-band OFDM link \cite{humphreys2023signal} with 15--60 Gbps capacity \cite{starlink-capacity-official, satellite-capacity-2, satellite-capacity}  per Starlink satellite)
to serve a vast amount of users. 
Since each LEO satellite only has a finite coverage, a constellation of satellites is necessary to form a global Internet coverage.
\todo{introduce satellite constellation layout, \eg, Walker constellation, for later theoretical analysis?}

To offer ubiquitous network services, a small constellation with 10s of LEO satellites suffices for global coverage (\eg, 66 satellites in Iridium \cite{iridium-constellation}).
But a small constellation does not have sufficient capacity to meet vast global users' demands for high-speed Internet.
To this end, {\em mega-constellations} with 1,000s--10,000s of satellites have been under rapid deployments to expand the LEO network capacity:
With around 5,200 active LEO satellites, Starlink has provisioned a 165 Tbps total network capacity \cite{spacex-total-capacity} for 2.3 million users in 70 countries \cite{starlink-subscribers} to offer 25--220 Mbps data speed per user \cite{starlink-performance-spec}. 

Of course, the LEO network capacity does not always grow with the mega-constellation scale.
For example, due to radio link interference among satellites, the LEO network capacity would start to saturate or even decrease
if scaled to around 30,000 satellites \cite{jia2021uplink}, 
which is luckily still far beyond operational mega-constellations' sizes today. 
\add{Instead, we reveal a more critical limiting factor of the LEO network capacity that today's mega-constellations are already experiencing.}


\subsection{The Congested Low Earth Orbits}
\label{sec:back:congested-space}

The recent rocket-fast deployment of mega-constellations has congested low Earth orbits and jeopardized space safety.
To date, there have been around 32,680 pieces of space debris and 14,450 satellites \cite{debris-number} at about 27,000 km/h in LEOs. 
Every fast-moving satellite is at high risk of physical collisions with debris or other satellites. 
Once collided, the satellite can break up into 10s--1,000s more pieces of debris to worsen orbit congestions and stimulate more collisions.
This phenomenon, called Kessler syndrome \cite{kessler2010kessler}, has occurred in the past decade \cite{collision-example-yunhai,collision-example-Fengyun} and severely threatened Starlink's satellite safety \cite{starlink-semi-annual-2021-7,starlink-semi-annual-2021-12,starlink-semi-annual-2022-7,starlink-semi-annual-2022-12,starlink-semi-annual-2023-6,starlink-semi-annual-2023-12} and network service quality \cite{starlink-degraded-service-congestion}.
\subsection{The Automatic Collision Avoidance}
\label{sec:back:collision-avoidance}

\begin{figure}[t]
\centering
\includegraphics[width=0.95\columnwidth]{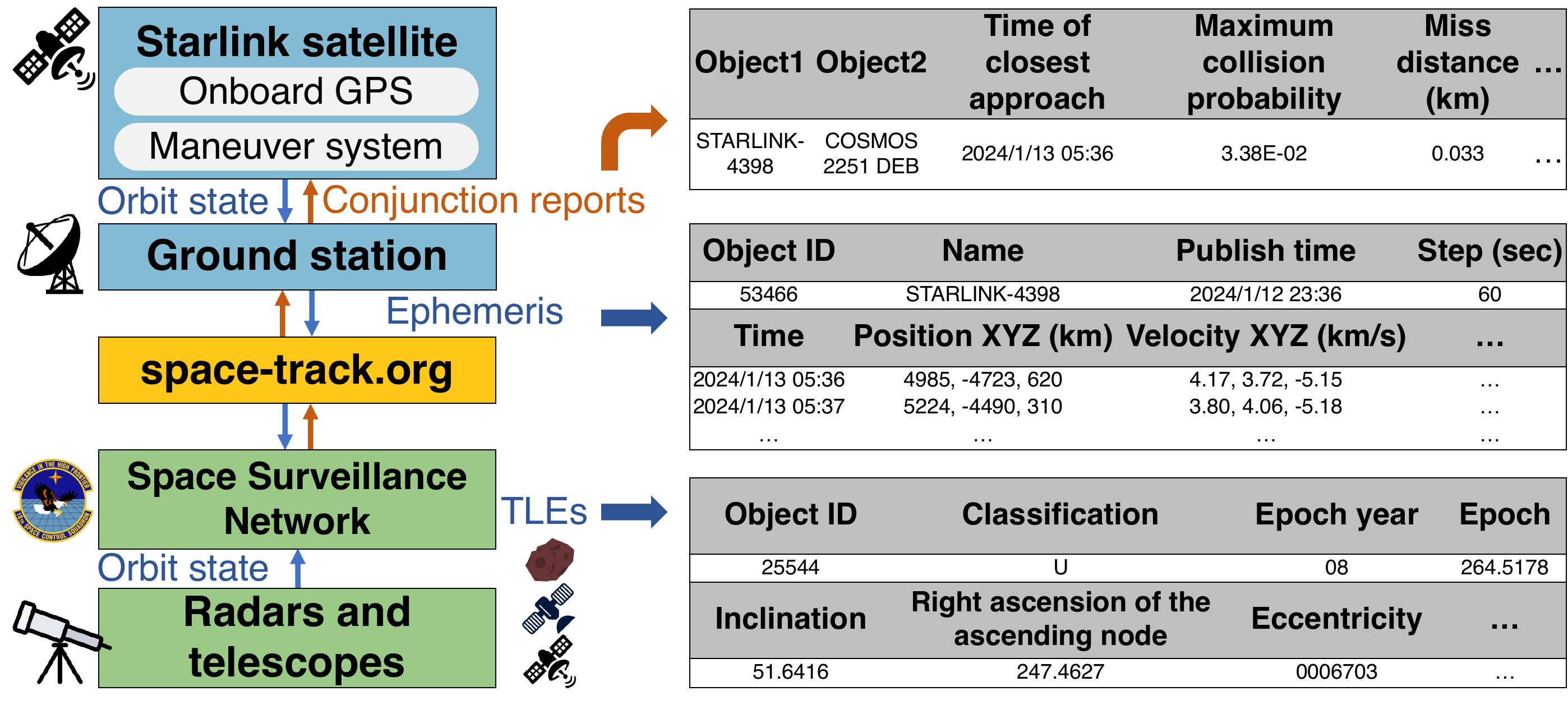}
\caption{Starlink's automatic collision avoidance \cite{starlink-maneuver,starlink-maneuver-2}.
}
\label{fig:starlink-ssa}
\end{figure}

To strive for LEO network infrastructure safety in congested orbits, satellite operators should forecast collision risks and instruct their satellites to maneuver for collision avoidance. 
They track their satellites' orbital motions using onboard GPS or terrestrial radars  \cite{sharma2000space,utzmann2014space,SBSS,maskell2008sapphire},
publicize them to the U.S. Space Surveillance Network (SSN) as fine-grained ephemeris or coarse-grained two-line elements (TLEs) \cite{tle} in Figure~\ref{fig:starlink-ssa}, 
and pull other debris and satellites' orbital traces from SSN.
With these data, they follow the method in Appendix~\ref{appendix:pc} to assess the pairwise collision risk 
between their satellite and every other space object (measured in the miss distance or collision probability $P_c$).
This forms pairwise conjunction reports \cite{ccsds-cdm,krage2020nasa} in Figure~\ref{fig:starlink-ssa} exchanged between operators.

If a satellite's most risky upcoming conjunction with another object exceeds its predefined safety threshold  ($P_c\geq10^{-4}$ in NASA \cite{krage2020nasa} and more stringent $P_c\geq10^{-5}$ in Starlink \cite{starlink-maneuver}), it should maneuver for collision avoidance. 
The common method is to slightly adjust its velocity using its propulsion system to avoid concurrent arrivals at the orbit intersection with the other object (analogous to acceleration/deceleration in terrestrial driving).
This method is energy-efficient and keeps this satellite in its legacy orbit for mission integrity \cite{sanchez2006collision}.
In rare cases when this conjunction will happen soon, the satellite may run emergency maneuvers by quickly raising/lowering its orbit for collision avoidance (analogous to lane change in terrestrial driving), which is uncommon due to its more fuel costs and mission interruptions \cite{mobicom23li,leonet23zhao}.
\todo{highlight pairwise collision avoidance, which is a common practice since the 1970s.}

Before LEO mega-constellations emerged, most satellite operators {manually} planned maneuvers at the remote control center. 
This method is not scalable or responsive to mega-constellations with many fast-moving LEO satellites.
Instead, recent constellations like Starlink \cite{starlink-maneuver, starlink-maneuver-2} and NASA Starling \cite{nasa-starling,probe2022prototype,brown2014simulated} have installed automatic maneuver systems in their satellites (Figure~\ref{fig:starlink-ssa}).
These systems 
have been proven successful in scaling to mega-constellations: As officially reported in \cite{starlink-semi-annual-2021-7,starlink-semi-annual-2021-12,starlink-semi-annual-2022-7,starlink-semi-annual-2022-12,starlink-semi-annual-2023-6,starlink-semi-annual-2023-12} and plotted in Figure~\ref{fig:semi-report_a}, Starlink has performed 
\add{10,000}s of maneuvers {\em every half year}. 

\section{Cascaded Collision Avoidance}
\label{sec:motivation}

While self-driving mega-constellations 
excel in avoiding collisions with {\em external} satellites/debris, 
we find their current maneuver paradigm can also amplify collision risks via {\em internal} {cascaded maneuvers}. 
This forces a dilemma between the LEO network's long-term lifetime and short-term capacity.
We introduce this phenomenon ($\S$\ref{sec:motivation:example}), 
analyze its root cause ($\S$\ref{sec:motivation:root-cause}), 
and validate its existence/popularity in Starlink ($\S$\ref{sec:motivation:validation}).

\subsection{An Illustrative Example}
\label{subsec:motivation-example}
\label{sec:motivation:example}

\begin{figure*}[t]
	\centering
	\vspace{-14mm}
	\includegraphics[width=\textwidth]{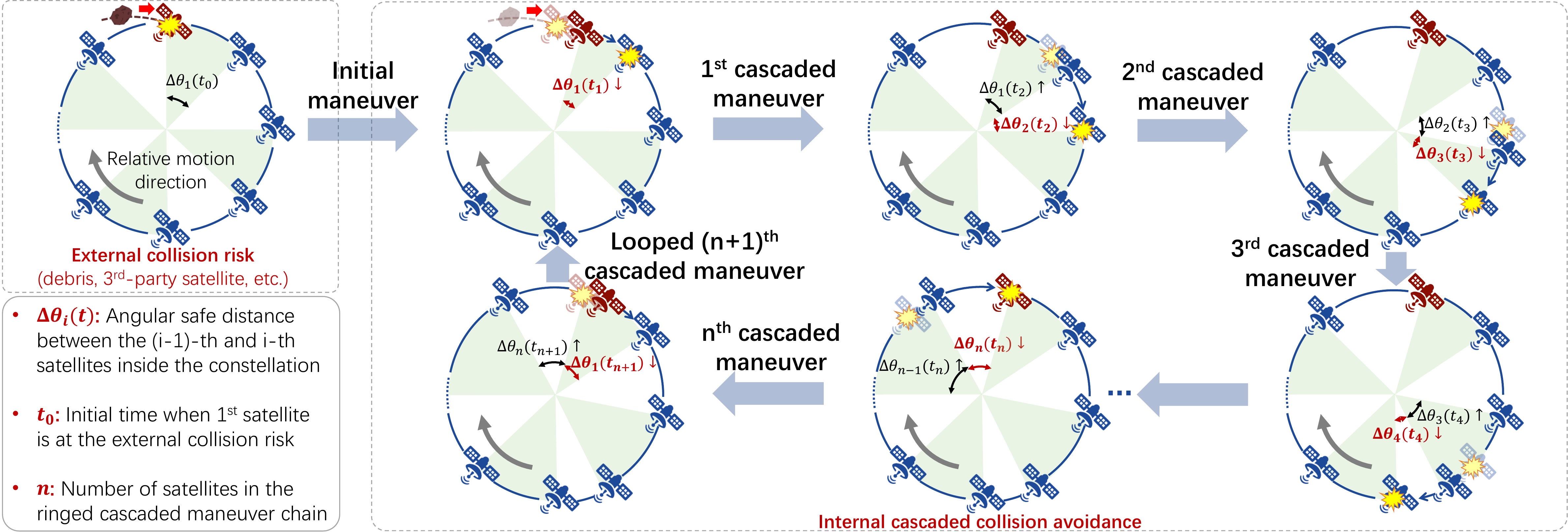}	
	\caption{An example of cascaded collision avoidance in a self-driving mega-constellation (online video at \cite{cascaded-maneuver-annimation}).}
	
	\label{fig:cascade_showcase}
	\vspace{-3mm}
\end{figure*}

Figure~\ref{fig:cascade_showcase} exemplifies the cascaded collision avoidance in a self-driving satellite mega-constellation's orbital shell. 
LEO satellites in this orbital shell move at a homogeneous velocity (measured in the angular mean motion \cite{mean-motion}) due to their identical orbital parameters.
To avoid collisions, they have been separated by an appropriate safe distance (measured in their relative phase differences $\Delta\theta_i$ in the orbital shell) during the mega-constellation design phase \cite{starlink-maneuver-2}.
Therefore, despite their fast mobility, LEO satellites inside the mega-constellation are not supposed to collide ideally. 

Now, consider the situation when one of these satellites (marked red in Figure~\ref{fig:cascade_showcase}) is approached by a 3rd-party satellite or external debris, which is common in congested LEOs ($\S$\ref{sec:back:congested-space}).
To avoid collisions, 
this satellite often accelerates/decelerates in orbit
to prevent concurrent arrivals at the intersection with the external object ($\S$\ref{sec:back:collision-avoidance}). 
In reality, most of these maneuvers have minimized their velocity and altitude change (typically 0.2--1 km \cite{fernandez2021impact}) to save satellites' energy \cite{probe2022prototype,brown2014simulated,sanchez2006collision}, thus seemingly harmless to the mega-constellation and network.

However, this maneuver for {\em external} collision avoidance can trigger cascaded {\em internal} collision risks.
During the above acceleration/deceleration, the red satellite in Figure~\ref{fig:cascade_showcase} also incurs relative motions to its neighboring satellites inside the mega-constellation.
This relative motion shortens its intra-constellation safe distance $\Delta\theta_1$ and raises collision risks with its neighbor. 
To this end, the neighboring satellite may also locally maneuver to enlarge its safe distance $\Delta\theta_1$, which, however, 
incurs further relative motions to later satellites and shortens their safe distance $\Delta\theta_2$.
This effect will propagate across satellites hop by hop inside the mega-constellation and cause cascaded maneuvers in Figure~\ref{fig:cascade_showcase}.
Since most satellites tend to adopt aggressive collision avoidance criteria 
for their own safety
($\S$\ref{sec:back:collision-avoidance}),
a small change of intra-constellation safe distance $\Delta\theta_i$ can easily satisfy these criteria and trigger cascaded maneuvers.
The denser satellites a mega-constellation comprises, the shorter the safe distance and the more maneuver propagation hops these satellites will have. 
So, the more frequent and exhaustive these cascaded maneuvers will be. 

Cascaded maneuvers are undesirable to the LEO mega-constellation network in various aspects: 

(1) Given limited fuels, each satellite can only perform finite collision avoidance maneuvers
(\eg, 350 times in Starlink \cite{starlink-maneuver-lifetime}).
Afterward, it will be permanently deorbited for space safety \cite{starlink-maneuver, starlink-maneuver-2}.
Cascaded maneuvers exhaust all satellites' maneuver budgets and shorten the LEO network lifetime.

(2) Most satellites cannot provide network services during maneuvers.
A maneuvering satellite's orientation change causes its antennas to temporarily lose alignment with the ground stations or terminals \cite{kepler-gsl-disruption-in-maneuver}.
Similarly, maneuvers can also disrupt optical inter-satellite links via out-of-alignments \cite{mobicom23li}.
In reality, each maneuver-induced satellite connectivity disruption can last for hours \cite{peiro2000galileo,esa-gsl-disruption-in-maneuver}. 
Cascaded maneuvers amplify this connectivity disruption's frequency.

(3) Frequent maneuvers can distort nominal satellite distributions in the mega-constellation, 
a well-known threat to GPS navigation \cite{peiro2000galileo,fan2017ground} and Earth observation \cite{aorpimai2007repeat,kim2021maintaining} missions.
In our context, cascaded maneuvers amplify space congestions in Figure~\ref{fig:cascade_showcase},
which is a cause of degraded network service experienced by Starlink user terminals \cite{starlink-degraded-service-congestion}.


\subsection{The Root Cause}
\label{sec:motivation:root-cause}

\p{Highlight {\bf ''vehicle-following'' model} in SOTA self-driving mega-constellations: not equal to ``local pairwise''.}

The root cause of cascaded maneuvers lies in each satellite's {\em local pairwise} collision avoidance.
As a decades-old paradigm starting from NASA \cite{krage2020nasa, operator-handbook, alfano2005numerical,probe2022prototype,brown2014simulated}, 
each satellite only assesses its own upcoming collision risks and maneuvers locally. 
As shown in $\S$\ref{sec:motivation:example}, 
while this local action is reasonable and cost-effective for each satellite, its
overall effects become undesirable for the global mega-constellation and network. 

\begin{figure}[t]
\centering
\includegraphics[width=\columnwidth]{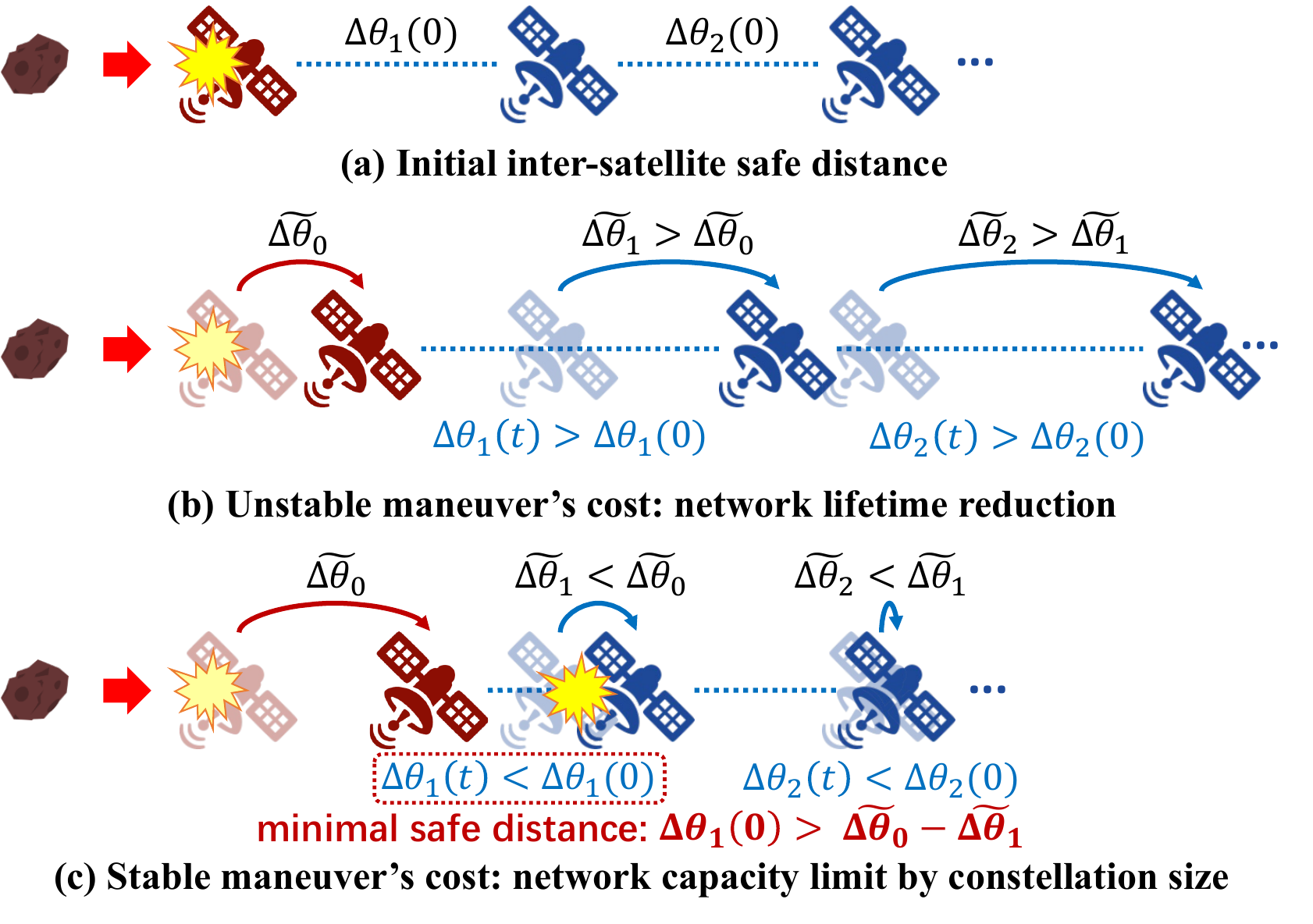}
\vspace{-5mm}
\caption{The LEO network lifetime-capacity dilemma due to cascaded internal collision avoidance.}
\label{fig:instability-cost}
\label{fig:stability-cost}
\label{fig:lifetime-vs-capacity}
\end{figure}

From the control system perspective, 
this phenomenon can be viewed as an unstable cascaded control. 
As shown in Figure~\ref{fig:lifetime-vs-capacity},
each satellite's local maneuver is a function that outputs its orbital motion deceleration (hence velocity/altitude change) in response to its safe distance and relative velocity to its closest neighboring satellite. 
Its deceleration output will update all neighboring satellites' safe distances (\ie, the input of their local maneuver policies), thus forming a cascaded transfer. 
If each transfer slightly amplifies the next satellite's safe distance (Figure~\ref{fig:lifetime-vs-capacity}b), 
then any small maneuver by external debris/satellite will be propagated and amplified hop by hop. 
As we will empirically show in $\S$\ref{sec:motivation:validation} and $\S$\ref{sec:validation}, this is common in reality because most satellites tend to enlarge their safe distances to prioritize their own safety.

To stop this undesirable amplification effect, an obvious fix seems to force each satellite's local maneuver to output a smaller deceleration compared to its inputs.
But as shown in Figure~\ref{fig:lifetime-vs-capacity}c, this fix raises internal collision risks since the inter-satellite spacing decreases hop by hop. 
To avoid collisions, the inter-satellite spacing inside the mega-constellation must be large enough to tolerate safe distance reductions.
This mandates sparser satellite distributions, thus limiting the mega-constellation scale and its LEO network capacity ($\S$\ref{sec:back:leo-network}).

Another possible solution is to shift to global maneuver coordination among all satellites 
using a central controller.
While seemingly effective, 
this option has two issues:
{\em (1) Scalability:} 
Coordinating all satellites' maneuvers
in a mega-constellation has been proven computationally intractable \cite{brown2014simulated}, which motivated distributed collision avoidance systems in the last decade \cite{starlink-maneuver, starlink-maneuver-2,nasa-starling,probe2022prototype};
{\em (2) Responsiveness:} 
The central controller at ground stations may not always be able to reach every fast-moving satellite for timely coordination.


\subsection{Real-World Instances in Starlink}
\label{subsec:motivation-validation}
\label{sec:motivation:validation}


\begin{figure*}[t]
\centering
\vspace{-14mm}
\subfloat[
By NASA's TROPICS satellite (amplification factor=22)
]{
\includegraphics[width=0.4\textwidth]{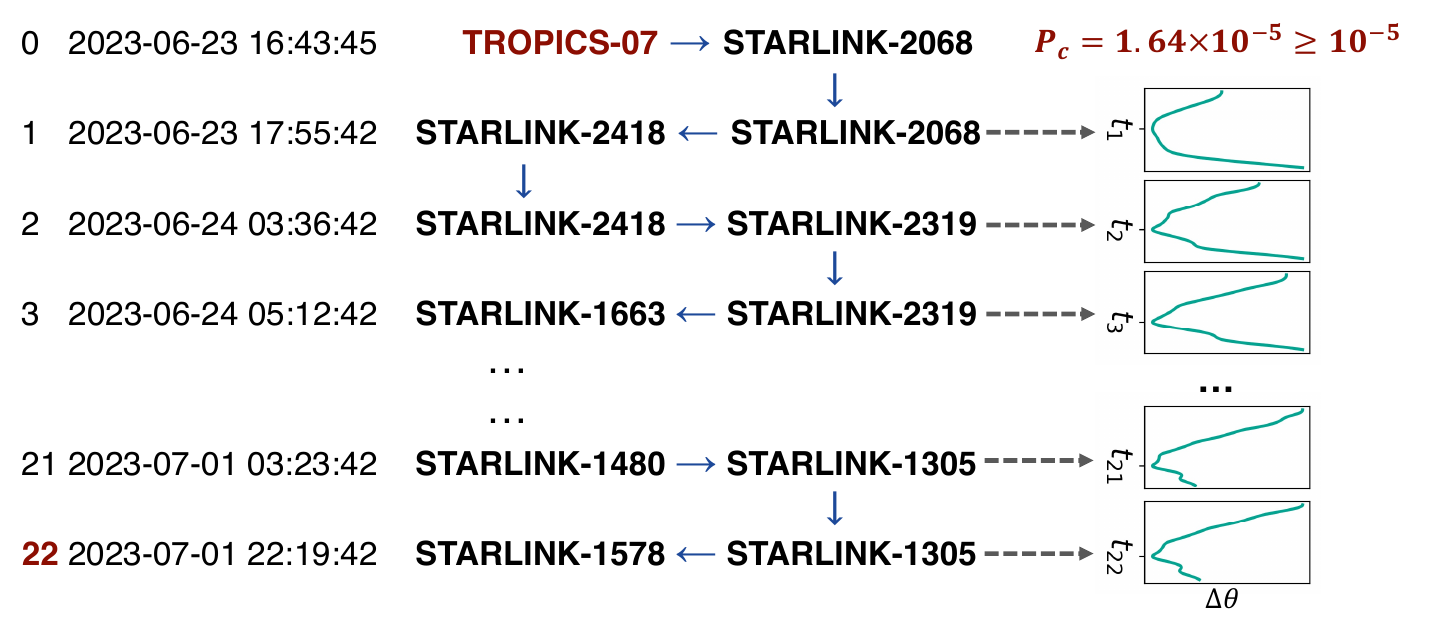}
\hspace{-3mm}
\includegraphics[width=0.2\textwidth]{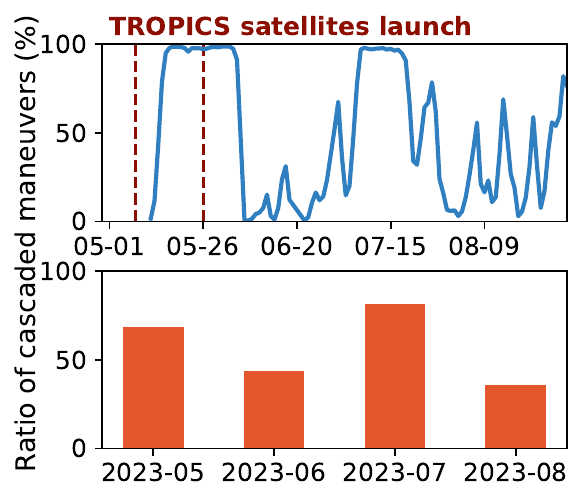}
\label{fig:tropics-showcase}
\label{fig:tropics-trend}
}
\subfloat[
By COSMOS 1408 debris  (amplification factor=27)
]{
\hspace{-3mm}
\includegraphics[width=0.4\textwidth]{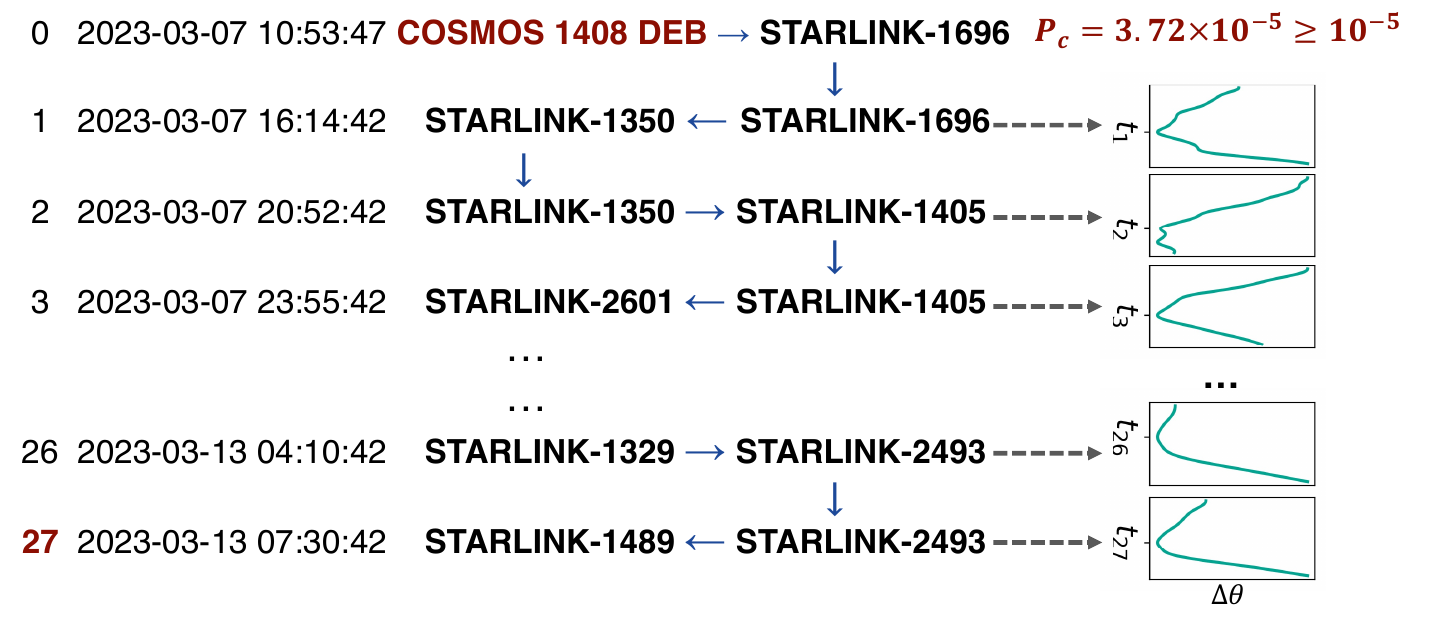}
\label{fig:cosmos-showcase}
}
\caption{Starlink's internal cascaded maneuvers in reality triggered by external 3rd-party satellites and debris.}
\label{fig:cascade-showcases}
\vspace{-4mm}
\end{figure*}

We find that cascaded collision avoidance indeed occurs at an alarming frequency in Starlink, the largest operational LEO satellite mega-constellation network. 
This section showcases it using our space situational awareness datasets in $\S$\ref{sec:validation:setup}. 
We leave more systematic empirical validations in $\S$\ref{sec:validation}.

\paragraphb{Existence of cascaded maneuvers:}
Figure~\ref{fig:cascade-showcases} showcases two real categories of cascaded collision avoidance in Starlink:

{\em $\circ$ Triggered by 3rd-party satellites:}
In May 2023, NASA launched its TROPICS satellites \cite{nasa-tropics} 
that operate at the same altitude as Starlink's satellites (\ie, 550 km). 
As shown in Figure~\ref{fig:tropics-showcase}, the TROPICS-07 satellite approached the Starlink-2068 satellite at a collision probability $P_c=1.64\times10^{-5}>10^{-5}$ (\ie, Starlink's threshold  \cite{starlink-maneuver}) on 2023/06, thus forcing Starlink-2068 to maneuver for collision avoidance,
shorten Starlink-2068's safe distance $\Delta\theta$ to its neighbor Starlink-2418,
and stimulate cascaded maneuvers. 
It lasts 22 hops until the cumulative altitude difference between the last-hop Starlink-1578 and the Starlink-1305 satellite is large enough to separate them safely.
This phenomenon partially explains NASA's report that Starlink experienced a surge of internal conjunction events after launching TROPICS satellites \cite{nasa-unique-conjunction-reports}. 

{\em $\circ$ Triggered by external debris:}
Russia's anti-satellite weapon test using its COSMOS-1408 satellite in 2021 \cite{russia-asat-test} generated numerous pieces of debris
and caused considerable collision avoidance maneuvers in Starlink \cite{starlink-semi-annual-2021-7,starlink-semi-annual-2021-12,starlink-semi-annual-2022-7,starlink-semi-annual-2022-12,starlink-semi-annual-2023-6,starlink-semi-annual-2023-12}.
Our datasets confirm them and further unveil cascaded maneuvers triggered by debris. 
As exemplified in Figure~\ref{fig:cosmos-showcase}, an external collision avoidance with a single COSMOS-1408 debris piece can incur \add{27} additional collision avoidance maneuvers inside Starlink. 

\paragraphb{Frequency of cascaded maneuvers:}
Figure~\ref{fig:semi-reports} counts the collision avoidance maneuvers in Starlink over time from our datasets, whose trend is consistent with Starlink's official semi-annual reports \cite{starlink-semi-annual-2021-7,starlink-semi-annual-2021-12,starlink-semi-annual-2022-7,starlink-semi-annual-2022-12,starlink-semi-annual-2023-6,starlink-semi-annual-2023-12} and NASA's conjunction report statistics \cite{nasa-unique-conjunction-reports}. 
It shows that most high-risk conjunction events occur among Starlink's satellites, rather than between Starlink and external objects. 
Cascaded maneuvers contribute to most of these internal conjunctions:
Every external collision avoidance triggers \add{up to 41} additional maneuvers inside the Starlink mega-constellation.

\begin{figure*}[t]
\centering
\vspace{-11mm}
\includegraphics[width=\textwidth]{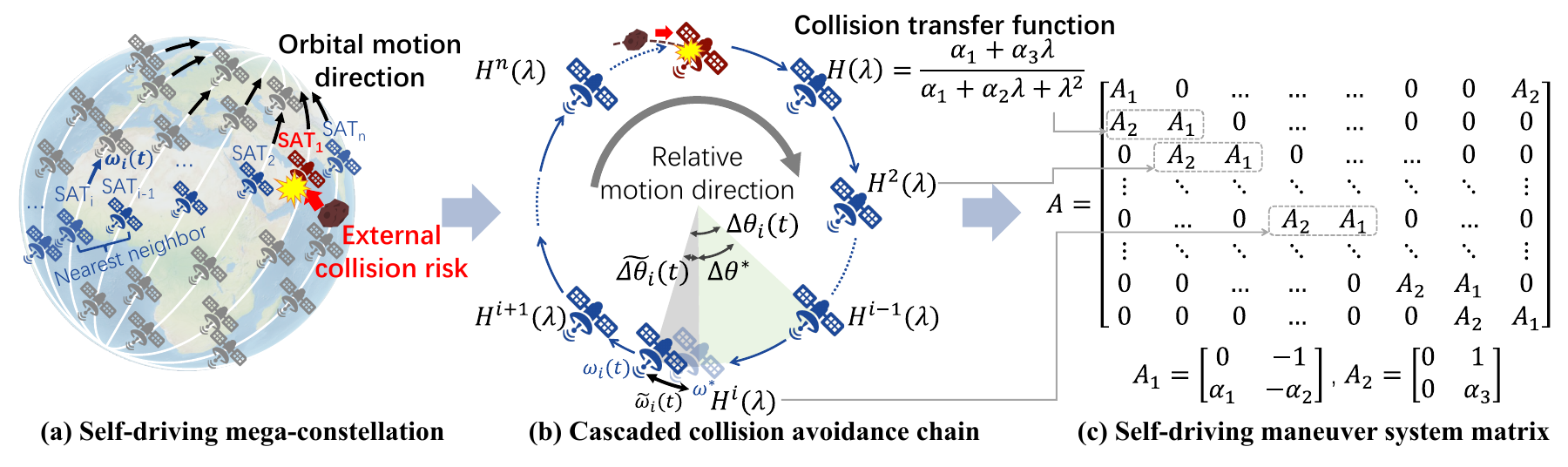}
\caption{Theoretical model of cascaded orbital collision avoidance in self-driving LEO satellite mega-constellations.}
\label{fig:theoretical-model}
\vspace{-4mm}
\end{figure*}

We further find that among these cascaded maneuvers, \add{97\%} of them occur between Starlink satellites from different orbits when approaching these orbits' intersections (exemplified in Figure~\ref{fig:theoretical-model}a).
Intra-orbit collision avoidance is infrequent due to more homogeneous satellite motions in the same orbit.

\begin{figure}[t]
\centering
\vspace{-5mm}
\subfloat[
Relative spacing between LEO satellites in adjacent orbits $F$ \cite{walker-constellation}
]{
\includegraphics[width=0.49\columnwidth]{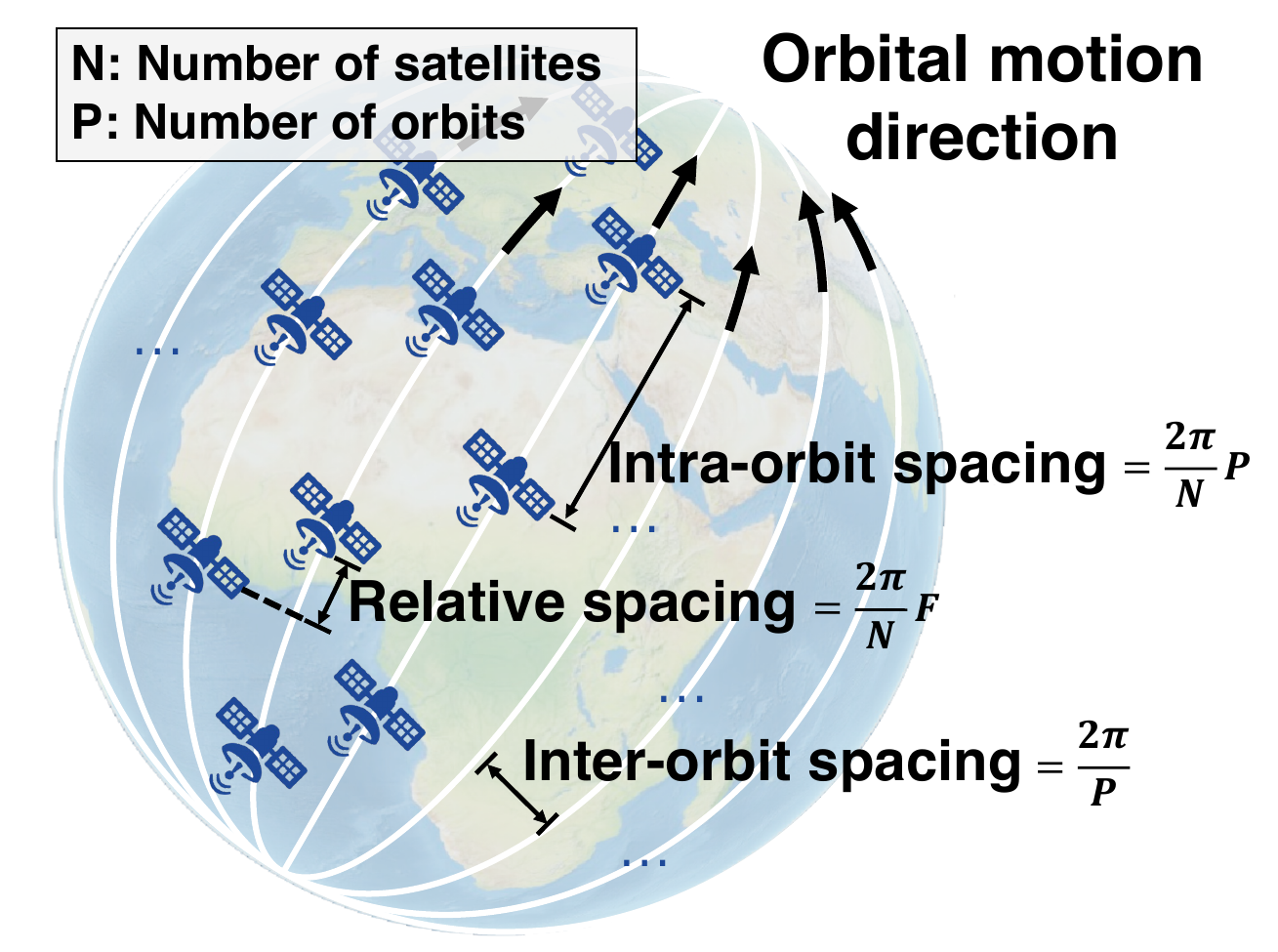}
\label{fig:orbit_para}
\label{fig:calculate_theta}
\label{fig:orbit_parameters}
}
\hfill
\subfloat[
The minimum inter-satellite safe distance under different $F$
]{
\includegraphics[width=0.45\columnwidth]{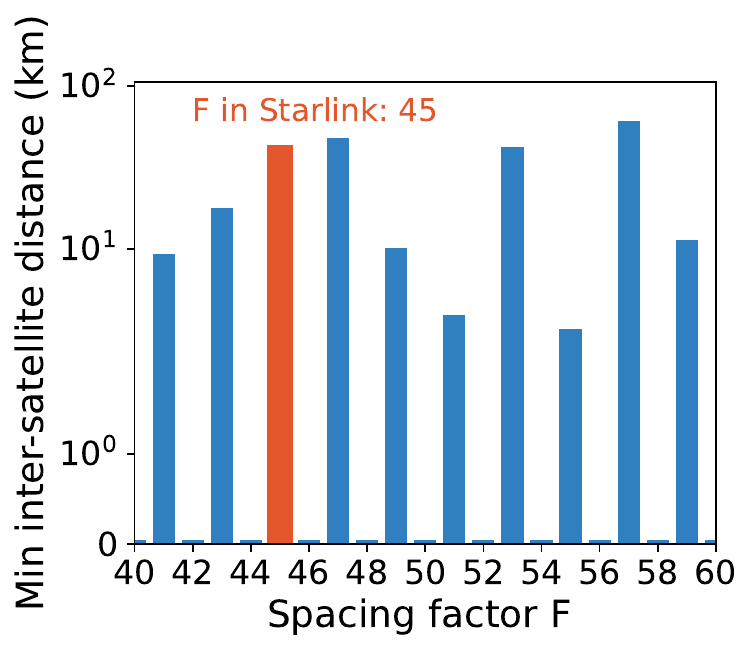}
\label{fig:F-value}
}


\caption{
Starlink has optimized inter-satellite spacing to mitigate collision risks inside its mega-constellation.
}
\label{fig:passive-deconflict}
\end{figure}

\paragraphb{Aren't they from satellite growth or constellation design defects?}
No.
Figure~\ref{fig:semi-report_a} shows Starlink's first-generation constellation's internal maneuvers continue growing even when its number of satellites in orbit remains almost unchanged from 2023/06 to 2023/12\footnote{During this period, Starlink is preparing its launches of second-generation mega-constellation and stops launching first-generation satellites \cite{starlink-launches}.}.
\add{We also empirically confirm Starlink's claim \cite{starlink-maneuver-2} that its constellation layout has been optimized to mitigate internal collisions. 
As shown in Figure~\ref{fig:passive-deconflict}, Starlink has chosen its spacing parameter $F$ in adjacent orbits \cite{walker-constellation} to enlarge the minimum safe distance between satellites. 
Despite this great effort, Figure~\ref{fig:semi-reports} shows that internal collision avoidance maneuvers are still common. 
}

\section{Self-Driving Stability Theory}
\label{sec:theory}

We next build a formal control-theoretic model for cascaded collision avoidance inside the self-driving mega-constellation ($\S$\ref{sec:theory:model}),
derive its stability condition ($\S$\ref{sec:theory:lyapunov-analysis}),
and characterize its fundamental dilemma between satellite network lifetime  ($\S$\ref{sec:theory:network-lifetime}) and capacity  ($\S$\ref{sec:theory:network-capacity}).

\subsection{Modeling Cascaded Collision Avoidance}
\label{subsec:model}
\label{sec:theory:model}

We consider a single orbital shell with $m$ identical orbits and $n$ identical satellites in the LEO mega-constellation, as shown in Figure~\ref{fig:theoretical-model}.
This model represents the most basic setting in operational mega-constellations today. 
In reality, satellites in the same orbital shell are manufactured and launched in batches, thus being homogeneous regarding their hardware, software, and maneuver policies. 
For LEO mega-constellations with multiple orbital shells at different altitudes and inclinations (\eg, Starlink~\cite{mobicom23li}), 
this basic model applies independently to each shell.
We next take four steps to model cascaded collision avoidance inside an orbital shell:

\paragraphb{(1) Collision risk from external debris/satellites:}
Each satellite in this orbital shell is surrounded by third-party satellites and external debris. 
To avoid collisions with them, it may conduct orbital maneuvers to deviate from its ideal position in the orbital shell, thus changing its relative positions to its neighboring satellites in the same orbital shell.

\paragraphb{(2) Each satellite's internal collision avoidance:}
The relative position change may raise the collision risk between neighboring satellites in the orbital shell, thus triggering internal collision avoidance maneuvers.
The state-of-the-art local pairwise collision avoidance paradigm in $\S$\ref{sec:back}--\ref{sec:motivation} finds each satellite's most threatening space object 
and instructs it to maneuver.
In the internal collision context, the most threatening object to each satellite inside the orbital shell is its closest neighboring satellite.
We thus model the $i$-th satellite's internal collision avoidance decision policy $F(\cdot)$ as
\begin{equation}
\dot{\omega}_i(t)=F(\Delta\theta_i(t),\dot{\Delta\theta}_i(t),\omega_i(t))
\label{eqn:self-driving-system}
\end{equation}
where $\Delta\theta_i(t)=\theta_{i-1}(t)-\theta_{i}(t)$ is this satellite's angular safe distance to its closest neighboring satellite  at time $t$, 
$\dot{\Delta\theta}_i(t)$ is the change of this angular safe distance, 
$\omega_i(t)$ is this satellite's orbital velocity (angular mean motion \cite{mean-motion}),
and $\dot{\omega}_i(t)$ is its angular acceleration for conducting internal collision avoidance maneuvers. 
We assume all satellites in the same orbital shell use the same internal collision avoidance policy $F(\cdot)$ due to their homogeneity. 
Our model allows for any general form of $F(\cdot)$ that satisfies two realistic assumptions:

\begin{compactenum}[\indent(a)]
\item For safety, each satellite's deceleration by its internal collision avoidance maneuver $\dot{\omega}_i(t)$ increases when its inter-satellite safe distance $\Delta\theta_i(t)$ decreases;
\item For safety, each satellite's deceleration by its internal collision avoidance maneuver 
$\dot{\omega}_i(t)$ increases when its preceding satellite's velocity $\omega_{i-1}(t)$ decreases.
\end{compactenum}

\paragraphb{(3) Cascaded collision avoidance chain:}
The above $(i-1)$-th satellite's maneuver changes its angular velocity ${\omega}_{i-1}(t)$ and relative distance to its closest succeeding neighboring satellite $i$ in the orbital shell, thus forcing this $i$-th satellite to conduct orbital maneuvers further using the policy $F(\cdot)$ in Equation~\ref{eqn:self-driving-system}.
In the circular orbital shell,
this results in a ringed cascaded collision avoidance chain in Figure~\ref{fig:theoretical-model}(b).
To determine this chain, we start from the first satellite that maneuvers for {\em external} collision avoidance,
find the second satellite in this chain as its closest successor regarding $\Delta\theta_1(t)$ inside this orbital shell,
iteratively repeat this process to find 
the $i$-th satellite as the $(i-1)$-th satellite's closest successor regarding $\Delta\theta_{i-1}(t)$, 
and stops after 
enumerating all satellites in this orbital shell 
or returning to the first satellite. 

\paragraphb{(4) Equilibrium state:}
To prevent wasting each satellite's lifetime, 
the cascaded maneuvers should eventually stop, 
\ie, $\lim_{t\rightarrow\infty}\dot{\omega}_i(t)=0$ and $\lim_{t\rightarrow\infty}\dot{\Delta\theta}_i(t)=0$ for all satellites $i$. 
At this equilibrium state, each satellite's orbital velocity $\omega^*$ and inter-satellite distance $\Delta\theta^*$ satisfy the following equation
\begin{equation}
F(\Delta\theta^*,0,\omega^*)=0
\label{eqn:equilibrium-point}
\end{equation}

Obviously, this equilibrium $x_e=(\Delta\theta^*,\omega^*)$ corresponds to the ideal satellite distribution in the constellation design. 
We aim to understand conditions of always converging to this equilibrium state 
and their impacts on satellite networking.
\subsection{Basic Lyapunov Stability Condition}
\label{subsec:condition}
\label{sec:theory:lyapunov-analysis}

To derive the conditions of converging to the original LEO constellation despite maneuvers, we leverage the Lyapunov stability analysis \cite{Lyapunov-stability} from the control theory. 
We note that the convergence to the equilibrium in Equation~\ref{eqn:equilibrium-point} is equivalent to ensuring the asymptotic Lyapunov stability as follows:

\begin{definition}[Lyapunov asymptotic stability \cite{Lyapunov-stability}]
The self-driving mega-constellation's equilibrium state $x_e=(\Delta\theta^*,\omega^*)$ in Equation~\ref{eqn:equilibrium-point} is saided to be asymptotically stable if for each $\epsilon>0,\exists\delta=\delta(\epsilon)>0$ so that if $||x(0)-x_e||<\delta$, then
\begin{eqnarray}
||x(t)-x_e||<\epsilon,\forall t\geq0\\ 
\lim_{t\rightarrow\infty}||x(t)-x_e||=0
\end{eqnarray}
\end{definition}

Intuitively, the Lyapunov asymptotic stability implies that all satellites will eventually return to the equilibrium state (\ie, the original constellation) despite any small perturbations due to external collision avoidance maneuvers.
We follow this formal definition to study the stability under small orbital perturbations by external maneuvers.
Consider a small perturbation of the $i$-th satellite from the equilibrium:
$$
\widetilde{\Delta\theta}_i(t)=\Delta\theta_i(t)-\Delta\theta^*,\:\:\:\:  \widetilde{\omega}_i(t)=\omega_i(t)-\omega^*
$$
Then we can apply the first-order Taylor series expansion to linearize this satellite's collision avoidance policy in Equation~\ref{eqn:self-driving-system} at the equilibrium state $x_e=(\Delta\theta^*,\omega^*)$ as
\begin{eqnarray}
\dot{\widetilde{\Delta\theta}}_i(t)&=&\omega_{i-1}(t)-\omega_i(t) = {\widetilde\omega}_{i-1}(t)-{\widetilde\omega}_i(t)\label{eqn:perturbation:theta}\\			
\dot{\widetilde{\omega}}_i(t)&=&\alpha_1 \widetilde{\Delta\theta}_i(t)-\alpha_2\widetilde{\omega}_i(t)+\alpha_3\widetilde{\omega}_{i-1}(t)\label{eqn:perturbation:omega}
\end{eqnarray}
where 
\begin{equation}
\alpha_1=\frac{\partial F}{\partial \Delta\theta}\vert_{x_e},\:\:\: \alpha_2=\left(\frac{\partial F}{\partial \dot{\Delta\theta}}-\frac{\partial F}{\partial \omega}\right)\vert_{x_e},\:\:\: \alpha_3=\frac{\partial F}{\partial \dot{\Delta\theta}}\vert_{x_e}
\label{eqn:self-driving-sensitivity}
\end{equation}
\ie, $\alpha_1$, $\alpha_2$, and $\alpha_3$ reflect each satellite's maneuver policy's sensitivity to the inter-satellite safe distance $\Delta\theta$, its own orbital motion velocity $\omega$, and inter-satellite relative velocity $\dot{\Delta\theta}$, respectively.
The assumptions (a) and (b) of the collision avoidance policy $F(\cdot)$ in $\S$\ref{sec:theory:model} implies $\alpha_1>0,\,\alpha_2>\alpha_3>0$.

Considering all $n$ satellites in each cascaded maneuver ring,
we can rewrite Equation~\ref{eqn:perturbation:theta}--\ref{eqn:perturbation:omega} 
in a matrix form.
Consider all satellites' small perturbations as a vector
$$
y(t)=[\widetilde{\Delta\theta}_1(t),\widetilde{\omega}_1(t),\widetilde{\Delta\theta}_2(t),\widetilde{\omega}_2(t),\cdots,\widetilde{\Delta\theta}_n(t),\widetilde{\omega}_n(t)]^T
$$
Applying this perturbation vector to Equation~\ref{eqn:perturbation:theta}--\ref{eqn:perturbation:omega} yields the global self-driving mega-constellation's system function
\begin{equation}
\dot{y}(t)=Ay(t)
\label{eqn:linear-self-driving-system}
\end{equation}
where $A$ is the $n\times n$ system matrix in Figure~\ref{fig:theoretical-model}c.
As a linear time-invariant system, this mega-constellation's self-driving system is asymptotically stable if and only if all of $A$'s complex-valued eigenvalues have negative real parts \cite{Lyapunov-stability}.
The following proposition reveals the general necessary and sufficient condition to achieve so (proved in Appendix~\ref{proof:stability-condition}):
\begin{prop}[Stability condition]
\label{prop:stability-condition}
The mega-constellation's maneuver policy in Equation~\ref{eqn:linear-self-driving-system} is asymptotically stable for any size $n$ of the self-driving mega-constellation if and only if
\begin{eqnarray}
\alpha_2^2-\alpha_3^2-2\alpha_1\geq0, \:\:\:\: \alpha_1>0, \:\:\:\: \alpha_2>\alpha_3>0
\end{eqnarray}
\end{prop}

\subsection{Instability's Cost: Network Lifetime}
\label{sec:theory:network-lifetime}
\label{subsec:cascade}


If a self-driving mega-constellation violates Proposition~\ref{prop:stability-condition},
its cascaded collision avoidance will exhaust each satellite's maneuver budgets and shorten the overall LEO network lifetime. 
We next quantify this
from two perspectives: 
the {\em microscopic} maneuver transfer between satellites, 
and the {\em macroscopic} maneuver growth trend in the mega-constellation.


\subsubsection{Microscopic View: Local Maneuver Amplification}

We first derive how a satellite's local maneuver is amplified hop by hop inside the mega-constellation.
Consider the $i$-th satellite's deviation from the equilibrium state in Equation~\ref{eqn:perturbation:theta}--\ref{eqn:perturbation:omega} due to its maneuver.
Since its system matrix $A$ is linear, this deviation can be always expressed as a linear combination of $A$'s exponential eigenfunctions \cite{Lyapunov-stability}\footnote{For a linear time-invariant system, its linear combination is equivalent to the discrete Fourier series \cite{oppenheim1999discrete} of each satellite's any small perturbation.}.
So, consider the $i$-the satellite's velocity deviation eigenfunction
\begin{equation}
\widetilde{\omega}_i(t)=c_ie^{\lambda t}
\label{eqn:eigenvector}
\end{equation}
where $\lambda$ is $A$'s eigenvalue and $e^{\lambda t}$ is its eigenfunction.
By substituting it into Equation~\ref{eqn:perturbation:theta}--\ref{eqn:perturbation:omega}, we get
\begin{equation}
\widetilde{\omega}_i(t)=H(\lambda)\widetilde{\omega}_{i-1}(t)
\label{eqn:collision-transfer}
\end{equation}
where
\begin{equation}
H(\lambda)=\frac{\alpha_1+\alpha_3\lambda}{\alpha_1+\alpha_2\lambda+\lambda^2}
\label{eqn:collision-transfer-function}
\end{equation}
is the {\bf collision transfer function} that propagates the leading satellite's motion ($\widetilde{\omega}_{i-1}$) to its follower's ($\widetilde{\omega}_{i}$).
As shown in Figure~\ref{fig:stability-cost}b, if $|H(\lambda)|\geq1$, each satellite's velocity and spacing deviation from the equilibrium state will be larger than its preceding satellite's, thus {amplifying cascaded maneuvers}.
In Appendix~\ref{proof:stability-condition}, we prove that $|H(\lambda)|\geq1$ is equivalent to the violation of the stability condition in Proposition~\ref{prop:stability-condition}. 

\subsubsection{Macroscopic View: Global Maneuver Growth Trend}

\begin{figure}[t]
\centering
\includegraphics[width=\columnwidth]{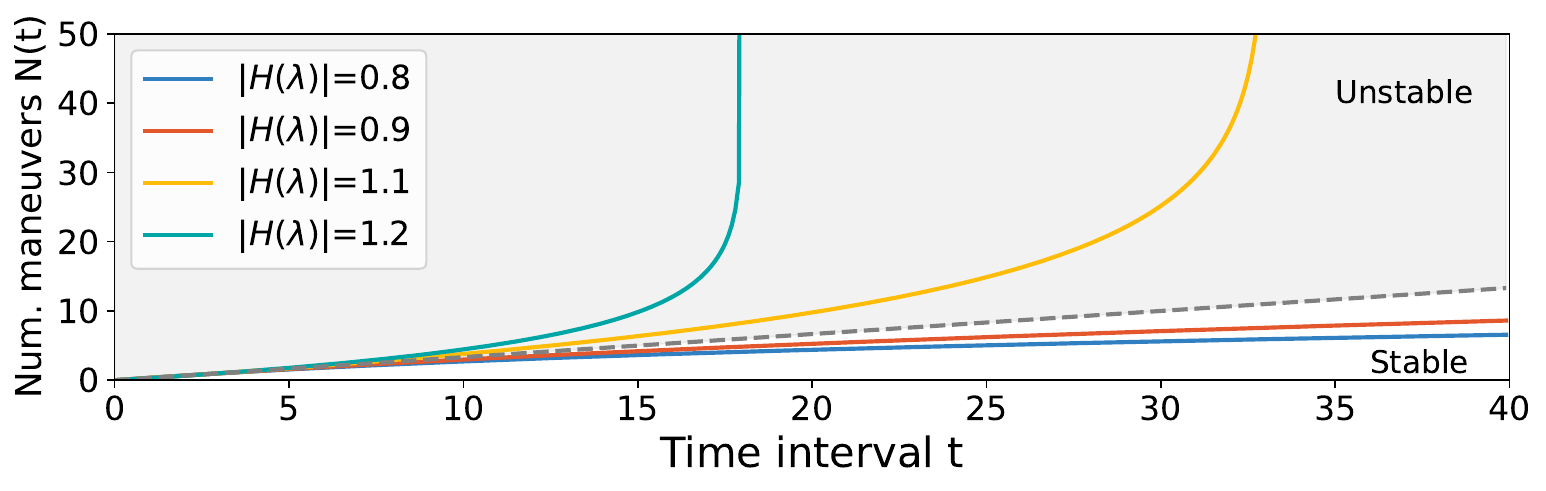}
\vspace{-4mm}
\caption{
Stability of self-driving mega-constellations under different orbital maneuver policies $H(\lambda)$. 
}
\label{fig:cascade_effect_theory}
\vspace{-2mm}
\end{figure}

We next switch to the global view to prove the {explosive} growth of orbital maneuvers when the mega-constellation's self-driving policy violates Proposition~\ref{prop:stability-condition}.
From Equation~\ref{eqn:collision-transfer-function}, the $i$-th satellite moves $|H(\lambda)|$ times faster than the $(i-1)$-th satellite due to its amplification effect in cascaded maneuvers.
Assume the uniformly distributed satellites in the mega-constellation.
If the first satellite takes time $t_0$ to approach the second satellite to trigger the first maneuver, 
then the $i$-th satellite would take $t_0/|H(\lambda)|^{i-1}$ time to approach the $(i+1)$-th satellite to trigger the $i$-th maneuver.
Summing them together,
we get the time to trigger the $N$-th maneuver:
\begin{equation}
t(N)=t_0\left(1-\frac{1}{|H(\lambda)|^n}\right)/\left(1-\frac{1}{|H(\lambda)|}\right)
\label{eqn:network-lifetime-t}
\end{equation}
Therefore, \add{given a limited maneuver budget $N$, the mega-constellation's lifetime $t$ decreases with a higher collision transfer function $|H(\lambda)|$. }
The total number of maneuvers $N(t)$ triggered by time $t$ can be derived from Equation~\ref{eqn:network-lifetime-t} as
\begin{equation}
N(t) = {\ln \left[ \frac{t_0}{t_0 - t (1 - |H(\lambda)|^{-1})}\right]}/{\ln |H(\lambda)|} 
\label{eqn:network-lifetime}
\end{equation}
As shown in Figure~\ref{fig:cascade_effect_theory}, the total number of maneuvers $N(t)$ grows explosively over time when the constellation's self-driving policy is unstable (\ie, $|H(\lambda)|\geq1$).
This trend is consistent with the observations from real space situational awareness datasets in Figure~\ref{fig:semi-reports} and Starlink's official reports \cite{starlink-semi-annual-2021-7,starlink-semi-annual-2021-12,starlink-semi-annual-2022-7,starlink-semi-annual-2022-12,starlink-semi-annual-2023-6,starlink-semi-annual-2023-12}. 
Unlike the common belief, this explosive maneuver growth can occur {\em without} adding more satellites in the mega-constellation.
It exhausts {\em all} satellite's finite maneuver budgets and dramatically shortens the entire LEO network's lifetime, as we will empirically quantify in $\S$\ref{sec:validation:lifetime}.

\subsection{Stability's Cost: Network Capacity}
\label{subsec:capacity-stability}
\label{sec:theory:network-capacity}

We next consider the scenario that the self-driving mega-constellation satisfies Proposition~\ref{prop:stability-condition} for stability.
In this case, it will not suffer from cascaded maneuvers in $\S$\ref{sec:theory:network-lifetime} since $|H(\lambda)|<1$, thus saving the LEO network's lifetime.
The cost, however, is the potential collision: 
To avoid maneuver amplifications, a satellite's maneuver's amplitude will be smaller than its preceding satellite's, thus reducing the safe distance between them (Figure~\ref{fig:instability-cost}c).
For collision avoidance, their safe distance before maneuvers should be sufficiently large to tolerate such damped maneuvers.
This limits the available satellites and total LEO network capacity ($\S$\ref{sec:back:leo-network}).

We now derive the upper bound of the mega-constellation size and LEO network capacity under the stable self-driving policy ($|H(\lambda)| < 1$).
In this case, 
Equation~\ref{eqn:collision-transfer-function} implies that the $(i-1)$-th satellite's acceleration amplitude is larger than the $i$-th satellite's due to cascaded maneuver, thus reducing their safe distance.  
As visualized in Figure~\ref{fig:stability-cost}c, to avoid collision between them, the $i$-th satellite's safe distance $\Delta\theta_i$ to the $(i-1)$-th satellite be no smaller than its accumulative reduction:
$$
\Delta\theta_i \geq \int_0^{\infty}\dot{\widetilde{\Delta\theta}}_i(t)dt = c_{i-1}\cdot\left[-\frac{1-H(\lambda)}{\lambda}\right],\forall i=1,2,\cdots,n
$$
according to Equation~\ref{eqn:perturbation:theta}, \ref{eqn:eigenvector} and \ref{eqn:collision-transfer-function}. 
Note that since the system matrix $A$ is stable, its $\lambda<0$ and collision transfer function $H(\lambda)<1$.
Considering all satellites along the cascaded ring and the worst-case maneuver amplitude $c_{max}$ to avoid external collisions, the minimum safe distance between any two satellites, denoted as $\Delta\theta_{safe}$, can be derived as 
\begin{equation}
\Delta\theta_{safe}>\max_i\Delta\theta_i=c_{max}\cdot \max_{\lambda}\left[-\frac{1-H(\lambda)}{\lambda}\right]
\label{eqn:safety-distance}
\end{equation}
To avoid internal collisions, 
this inter-satellite safe distance threshold directly limits the number of satellites in each mega-constellation.
Consider the Walker constellation in Figure~\ref{fig:orbit_para} used by operational LEO networks. 
Given its inter-satellite spacing parameter in adjacent orbits $F$, 
a self-driving constellation's maximum number of satellites $n$ (\ie, its total LEO network capacity according to $\S$\ref{sec:back:leo-network}) is limited by
\begin{equation}
n<2\pi\cdot F/\Delta\theta_{safe}
\label{eqn:leo-capacity}
\end{equation}
This network capacity bound decreases monotonically with a stabler self-driving policy. 
As we will empirically validate in $\S$\ref{sec:validation:capacity}, 
it is more stringent than the RF interference-induced LEO network capacity limit in $\S$\ref{sec:back:leo-network} \cite{jia2021uplink}, thus becoming the bottleneck to offer high-speed Internet access for more users.

\paragraphb{Satellite network lifetime-capacity dilemma:}
Our analysis has revealed a fundamental dilemma between the satellite network lifetime and capacity in state-of-the-art self-driving mega-constellations.
From Equation~\ref{eqn:collision-transfer} and Equation~\ref{eqn:safety-distance}, it is clear that a larger collision transfer function $|H(\lambda)|$ helps expand the satellite network capacity at the cost of more maneuvers and thus shorter network lifetime, and vice versa.
When the self-driving policy $|H(\lambda)|\rightarrow1$, the satellite network capacity is almost unleashed, but maneuvers will also be propagated hop by hop without convergence to shorten the satellite network lifetime.
No local pairwise maneuver parameters in Equation~\ref{eqn:self-driving-sensitivity} exist to work around this dilemma.


\section{Experimental Validation}
\label{sec:validation}

We validate our theory with Starlink's large-scale space situational awareness datasets. 
We infer its proprietary maneuver policy from real data ($\S$\ref{sec:validation:setup}),
cross-check its stability
using our theory and actual cascaded maneuvers ($\S$\ref{sec:validation:theory}),
and assess its impacts on network lifetime ($\S$\ref{sec:validation:lifetime}) and capacity ($\S$\ref{sec:validation:capacity}). 

\subsection{Experimental Setup}
\label{subsec:validation-setup}
\label{sec:validation:setup}

Our verification experiments over Starlink takes four steps:

\paragraphb{(I) Datasets:}
Table~\ref{tab:dataset} summarizes three categories of large-scale space situational awareness datasets in our study:

{\bf $\circ$ Fine-grained ephemeris:}
To date, Starlink is the only satellite network operator that volunteers to publicize its satellites' massive fine-grained orbital motion traces \cite{starlink-maneuver, starlink-maneuver-2}. 
As exemplified in Figure~\ref{fig:cascade_showcase} and summarized in Table~\ref{tab:dataset},
its unique ephemeris dataset records each satellite's runtime position, velocity, and their covariance from its onboard GPS on a minute basis. 
This offers sufficient granularity for us to reverse-engineer Starlink's self-driving maneuver policy.

{\bf $\circ$ Coarse-grained ephemeris (TLEs):}
Except for Starlink, other satellites and debris' fine-grained positions are not publicly available.
Instead, their coarse-grained orbital parameters are 
publicized by the U.S. SSN on an hourly basis in the TLE format in Figure~\ref{fig:starlink-ssa} \cite{tle}.
With TLEs, we can estimate each debris/satellite's location using the classic SGP4 orbit propagator \cite{hoots1980models} with a cumulative error of 1--3 km per day,
which suffices in this study to detect external collision risks as initial triggers for cascaded maneuvers.

{\bf $\circ$ Pairwise conjunction reports:}
They are standardized data \cite{ccsds-cdm,krage2020nasa} to guide collision avoidance today. 
Given any two space objects' position and velocity, they follow procedures in Appendix~\ref{appendix:pc} to estimate and report their collision probability to satellite operators for maneuver planning. 
We use conjunction reports to detect and correlate maneuvers.

\begin{table}[t]
\vspace{-9mm}
 \resizebox{\columnwidth}{!}{{
   \begin{tabular}{p{3cm}|c|c|c}
    \toprule
     {\bf Dataset type} & {\bf Starlink's} & {\bf Two-Line}  & {\bf Conjunction} \\
     	& {\bf  ephemeris} &{\bf Elements (TLEs)}& {\bf reports}\\
    \hline
    {\bf Source} & \multicolumn{2}{c|}{space-track.org} & celestrak.org \\
    \hline
    {\bf Time span} & 2022/12--2024/01 & 2019/05--2024/01 & 2019/12--2024/01 \\
    \hline
    {\bf Num. records} & 14,397,368,913 & 55,209,626 & 16,799,257 \\
    \hline
    {\bf Record frequency} & 1 minute & 3.0--34.7 hrs & 8 hrs \\
    \hline
    {\bf Num. space objects} & 5,309 & 24,237  & 24,435\\
    \bottomrule    
   \end{tabular}
 }}
 \caption{
Space situational awareness datasets.
 } 
 \label{tab:dataset}
 \vspace{-2mm}
\end{table}

\paragraphb{(II) Inferring ``black-box'' maneuver policies:}
Starlink generally follows the local pairwise maneuver paradigm \cite{starlink-maneuver, starlink-maneuver-2}, but its detailed form of maneuver decision policy $F(\cdot)$ in Equation~\ref{eqn:self-driving-system} is proprietary.
Previous work 
\cite{mobicom23li} takes the first step in detecting its maneuvers by comparing a satellite's actual location with its sparse TLE-based prediction. 
But this heuristic maneuver detector cannot unveil sufficient information about $F(\cdot)$ from the coarse-grained TLEs to help evaluate its stability and impacts on satellite networking.


To this end, we develop a rigorous approach to unveil Starlink's maneuver policy from its fine-grained ephemeris.
Our key insight is that, while the exact closed-form expression of the maneuver policy $F(\cdot)$ is unknown, 
its sensitivities to each satellite's runtime velocity and inter-satellite spacing (both being observable from our ephemeris data)
can be directly derived by applying the differential Equation~\ref{eqn:self-driving-sensitivity}. 
Therefore, we extract Starlink's constellation design parameters from its public FCC fillings~\cite{starlink-constellation-layout-gen1},
deduce its equilibrium state $x_e$ from them,
track each satellite's runtime angular velocity $\omega$, inter-satellite spacing $\Delta\theta$, and maneuver acceleration $\dot{\omega}$ from its ephemeris,
calculate these parameters' deviations from $x_e$, 
and derive the maneuver policy's sensitivity parameters $(\alpha_1,\alpha_2,\alpha_3)$ using Equation~\ref{eqn:self-driving-sensitivity}.
These parameters $(\alpha_1,\alpha_2,\alpha_3)$ suffice to characterize Starlink's maneuver stability and impacts on networking via Proposition~\ref{prop:stability-condition} and Equation~\ref{eqn:collision-transfer-function}--\ref{eqn:leo-capacity}. 

\paragraphb{(III) Extracting real cascaded collision avoidance:}
To validate our theory in $\S$\ref{sec:theory}, we extract Starlink's cascaded maneuvers from our datasets as the ground truth as follows:

{\em (1) Detect the triggering maneuver by external collision risks:}
Since the external debris/satellites' fine-grained ephemeris is not publicly available,
we follow \cite{mobicom23li} to detect Starlink-on-others maneuvers based on their conjunction reports: 
A Starlink-on-others maneuver is detected if its collision probability 
$P_c\geq10^{-5}$  \cite{starlink-maneuver, starlink-maneuver-2}
and the Starlink satellite's real semi-major axis deviates from its prediction by $\geq$1 km.

{\em (2) Search subsequent internal maneuvers:} 
For each Starlink-on-others maneuver, we examine if it stimulates cascaded Starlink-on-Starlink maneuvers.
We leverage Starlink's fine-grained ephemeris to accurately check if the maneuvering satellite's safe distance $\Delta\theta$ to its internal neighbors decreases (indicating an increased collision risk).
If so, we further check if its neighbor later decelerates to increase this $\Delta\theta$ for collision avoidance, as exemplified in Figure~\ref{fig:cascade-showcases}.
If true, we mark it as a cascaded maneuver and iterate this process on this neighbor to extract the cascaded collision avoidance chain.

\paragraphb{(IV) Assessing maneuvers' impacts on networking:}
For the LEO network lifetime, 
We follow Starlink's report \cite{starlink-maneuver-lifetime} to assume a 350-maneuver budget per satellite,
count cascaded maneuvers, 
compare them with the total maneuvers, 
and estimate LEO network lifetime reductions. 
For the LEO network capacity, we follow Starlink's official report \cite{satellite-capacity} to
assume a 20 Gbps capacity per satellite,
evaluate its total network capacity upper bound in Equation~\ref{eqn:safety-distance}--\ref{eqn:leo-capacity}, 
and assess how close Starlink's network capacity is to this bound.
\subsection{Instability of Real Maneuver Policies}
\label{sec:validation:theory}

\begin{figure}[t]
\centering
\vspace{-14mm}
\subfloat[
Maneuver policy over time
]{
\includegraphics[width=0.51\columnwidth]{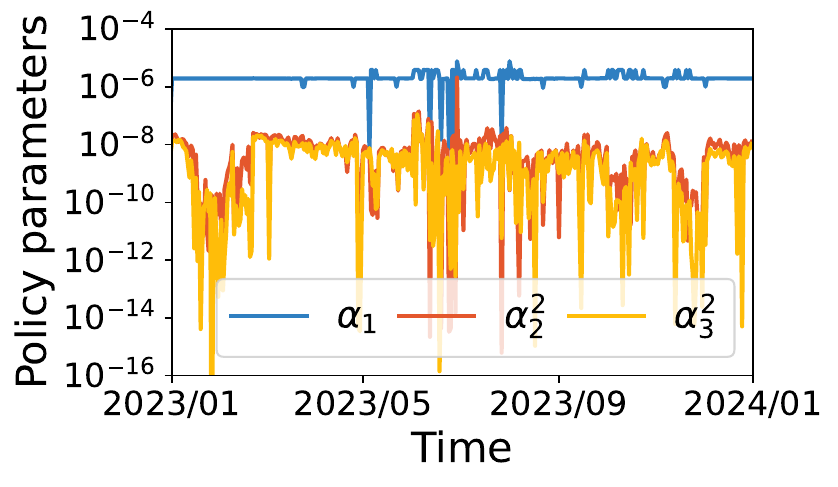}
\label{fig:param-over-time}
}
\subfloat[
Distribution of parameters
]{
\includegraphics[width=0.48\columnwidth]{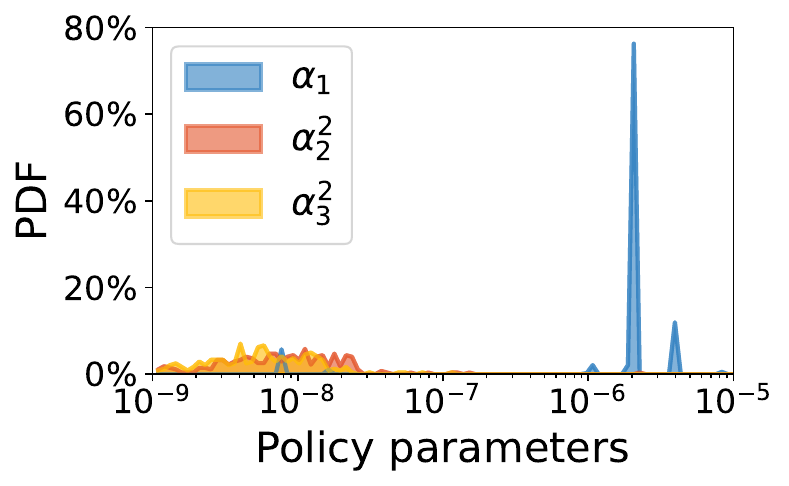}
\label{fig:dist-parameter}
}
\caption{
Starlink's local pairwise maneuver policy.
}
\label{fig:exp-policy-parameters}
\end{figure}
\begin{figure}[t]
\vspace{-8mm}
\centering
\subfloat[
Stability violation (Proposition~\ref{prop:stability-condition})
]{
\includegraphics[width=0.51\columnwidth]{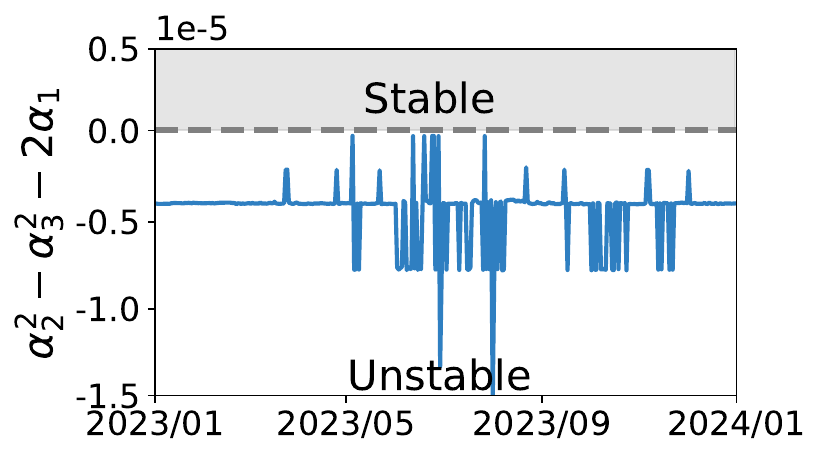}
\label{fig:exp-condition}
}
\subfloat[
Cascaded maneuver frequency
]{
\includegraphics[width=0.50\columnwidth]{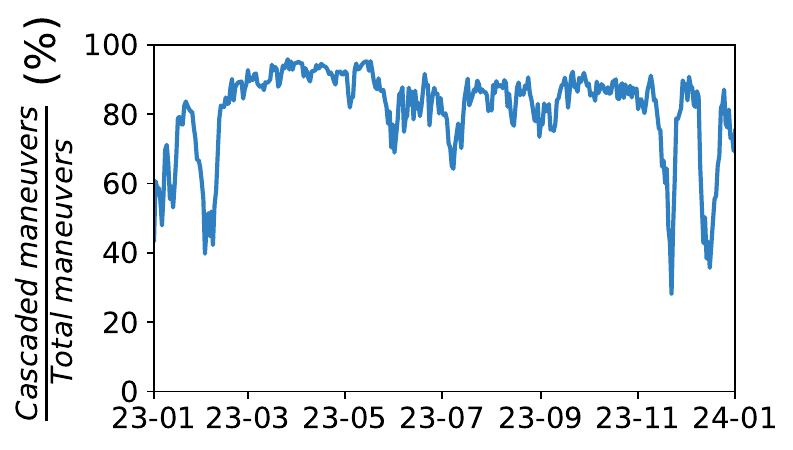}
\label{fig:cascaded-conjunction}
}
\caption{
Instability of Starlink's self-driving policy.
}
\label{fig:exp-stability-condition}
\vspace{-2mm}
\end{figure}

We first characterize Starlink's internal collision avoidance policy, assess its stability using Proposition~\ref{prop:stability-condition}, and compare this theoretical result with experimental validations.

\paragraphb{Characteristics of Starlink's maneuver policy:}
Figure~\ref{fig:exp-policy-parameters} demonstrates Starlink's maneuver policy's sensitivity to inter-satellite spacing ($\alpha_1$), satellite velocity ($\alpha_2$), and inter-satellite relative velocity ($\alpha_3$) in Equation~\ref{eqn:self-driving-sensitivity} extracted from our datasets using our method in $\S$\ref{sec:validation:setup}.
We make two observations:

First, Starlink's maneuver decision is more sensitive to inter-satellite spacing than velocity: $\alpha_1$ is two orders larger than $\alpha_2^2$ and $\alpha_3^2$ in Proposition~\ref{prop:stability-condition} most time.
This is consistent with most satellite operators' collision avoidance procedure in Appendix~\ref{appendix:pc} \cite{krage2020nasa, operator-handbook, alfano2005numerical},
in which the inter-satellite spacing (miss distance) dominates their collision probability calculation and maneuver decision.
The relative velocity between satellites only fine-tunes the collision probability calculation.

Second, Starlink's dominating maneuver parameter, inter-satellite distance sensitivity $\alpha_1$, remains generally stable except for an apparent adjustment in 2023.06--2023.09.
We gauge that this update may be attributed to Starlink's joint collision avoidance system test with NASA's experimental Starling satellites \cite{Starling-test,nasa-starling,probe2022prototype} in 2023.07.
Other maneuver parameters about satellite velocity also conform to common assumptions of maneuver policies $\alpha_2>\alpha_3>0$ in $\S$\ref{sec:theory:model}--\ref{sec:theory:lyapunov-analysis} most of the time.
Rare exceptions exist due to fluctuations of elliptical orbital parameters but are tolerable in our theory. 

\paragraphb{Theoretical vs. experimental stability analysis:}
As shown in Figure~\ref{fig:exp-condition},
Starlink's maneuver policy in Figure~\ref{fig:exp-policy-parameters} has consistently violated the stability condition in Proposition~\ref{prop:stability-condition}.
This theoretical violation conforms to our empirical results in Figure~\ref{fig:cascaded-conjunction} that on average, \add{81.4\%} of Starlink-on-Starlink maneuvers belong to the cascaded collision avoidance chain inferred using our method in $\S$\ref{sec:validation:setup}.
It explains the explosive growth of maneuvers in Figure~\ref{fig:semi-report_a} and $\S$\ref{sec:motivation:validation} despite Starlink's excellent mega-constellation design.
As shown in Figure~\ref{fig:exp-policy-parameters}, this stability violation mainly stems from Starlink's maneuver policy's high sensitivity to inter-satellite spacing $\alpha_1$. 

\subsection{Impacts on LEO Network Lifetime}
\label{sec:validation:lifetime}

\begin{figure}[t]
\centering
\vspace{-12mm}
\subfloat[
Remaining network lifetime
]{
\includegraphics[width=0.46\columnwidth]{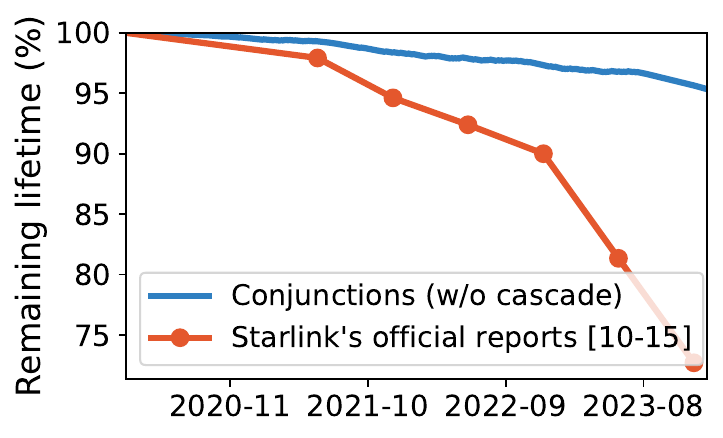}
\label{fig:impact-lifetime}
}
\subfloat[
Decayed satellites
]{
\includegraphics[width=0.465\columnwidth]{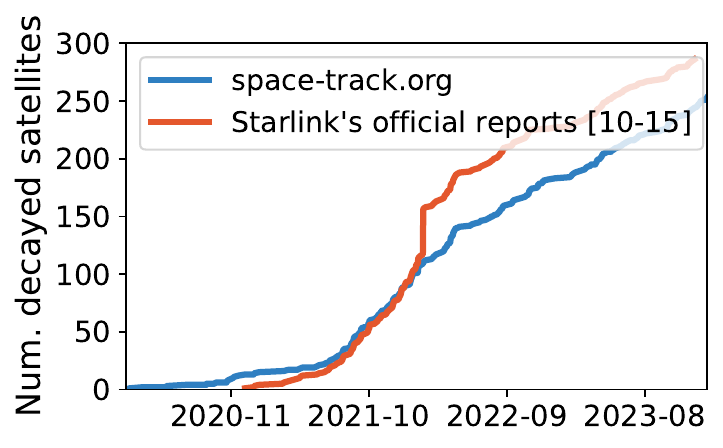}
\label{fig:impact-ratio-operational-sat}
\label{fig:decayed-satellite}
}
\vspace{-4mm}
\subfloat[
Maneuver amplification factor in orbital shell 1 and 2
]{
\includegraphics[width=0.55\columnwidth]{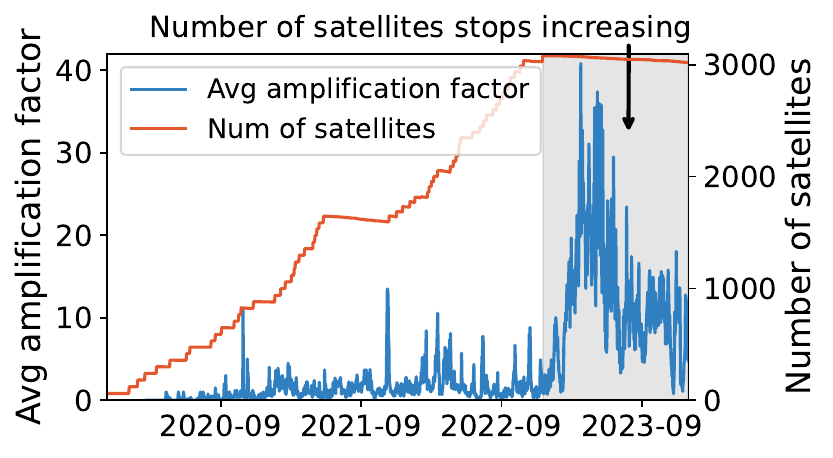}
\includegraphics[width=0.388\columnwidth]{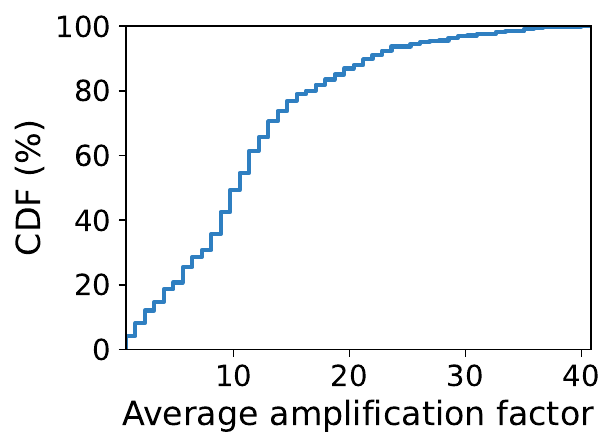}
\label{fig:impact-amplification-factor-cdf}
\label{fig:impact-amplification-factor}
}
\caption{
Starlink's network lifetime reduction.
}
\label{fig:impact-cascade-maneuver-lifetime}
\vspace{-1mm}
\end{figure}

We next evaluate how Starlink's unstable maneuver policy in $\S$\ref{sec:validation:theory} threatens its network lifetime. 
Figure~\ref{fig:cascaded-conjunction} shows that cascaded maneuvers by Starlink's unstable policy contribute \add{81.4\%} of its total maneuvers.
They exhaust each satellite's maneuver budgets and shorten Starlink's network lifetime.
As shown in Figure~\ref{fig:impact-lifetime}, assuming Starlink's 350-maneuver budget per satellite \cite{starlink-maneuver-lifetime}, 
cascaded maneuvers have wasted \add{79.5\%} of Starlink's total network lifetime.
\add{This trend is consistent with Starlink's non-trivial satellite removal from orbits based on its official reports \cite{starlink-semi-annual-2021-7,starlink-semi-annual-2021-12,starlink-semi-annual-2022-7,starlink-semi-annual-2022-12,starlink-semi-annual-2023-6,starlink-semi-annual-2023-12} and our data in Figure~\ref{fig:decayed-satellite}.}

In theory, an unstable policy would lead to unstoppable cascaded maneuvers (Equation~\ref{eqn:network-lifetime}). 
In practice, cascaded maneuvers will eventually stop
since the motion amplification in Equation~\ref{eqn:collision-transfer}--\ref{eqn:collision-transfer-function} will lower the satellite altitude based on Kepler's third law.
After sufficient hops of amplification, a satellite will eventually greatly deviate from its original altitude to avoid collisions with its neighbors and stop maneuver propagations.
The exact number of hops to stop cascaded maneuvers depends on various factors, such as the satellite size, altitude, and initial maneuver magnitude.
We define the {\em maneuver amplification factor} as the number of additional maneuvers triggered by an external collision avoidance. 
Figure~\ref{fig:impact-amplification-factor} shows Starlink's maneuver amplification factor fluctuates \add{from 1 to 41} over time.
It even grows when its number of satellites remains stable due to the co-orbit TROPICS satellite deployment in Figure~\ref{fig:tropics-showcase} and $\S$\ref{sec:motivation:validation}.
\fixme{Besides, cascaded maneuvers can also be interrupted by additional external maneuvers, thus affecrting amplification factors.}

\subsection{Impacts on LEO Network Capacity}
\label{sec:validation:capacity}

We last assess the potential reduction of network capacity if Starlink stabilized its maneuver policy in $\S$\ref{sec:validation:theory}.
Since its maneuver policy is dominated by the inter-satellite spacing and relative motion ($\S$\ref{sec:validation:theory}),
we consider the feasibility of fine-tuning its sensitivity to both factors (\ie, $\alpha_1$ and $\alpha_3$) to satisfy Proposition~\ref{prop:stability-condition} while maximizing the LEO network capacity in Equation~\ref{eqn:safety-distance}--\ref{eqn:leo-capacity}.
We compare this theoretical LEO network capacity bound with Starlink's current capacity and characterize its tradeoff with the LEO network lifetime.

\begin{figure}[t]
	\centering
	\vspace{-12mm}
	\subfloat[The size of mega-constellation]{
	\includegraphics[width=0.5\columnwidth]{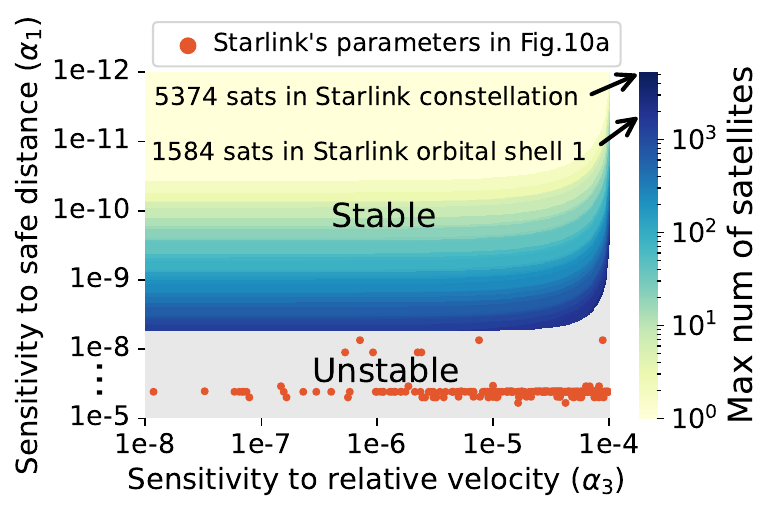}
	\label{fig:experimental-capacity-a}
	}
	\subfloat[Network capacity vs lifetime]{
	\includegraphics[width=0.47\columnwidth]{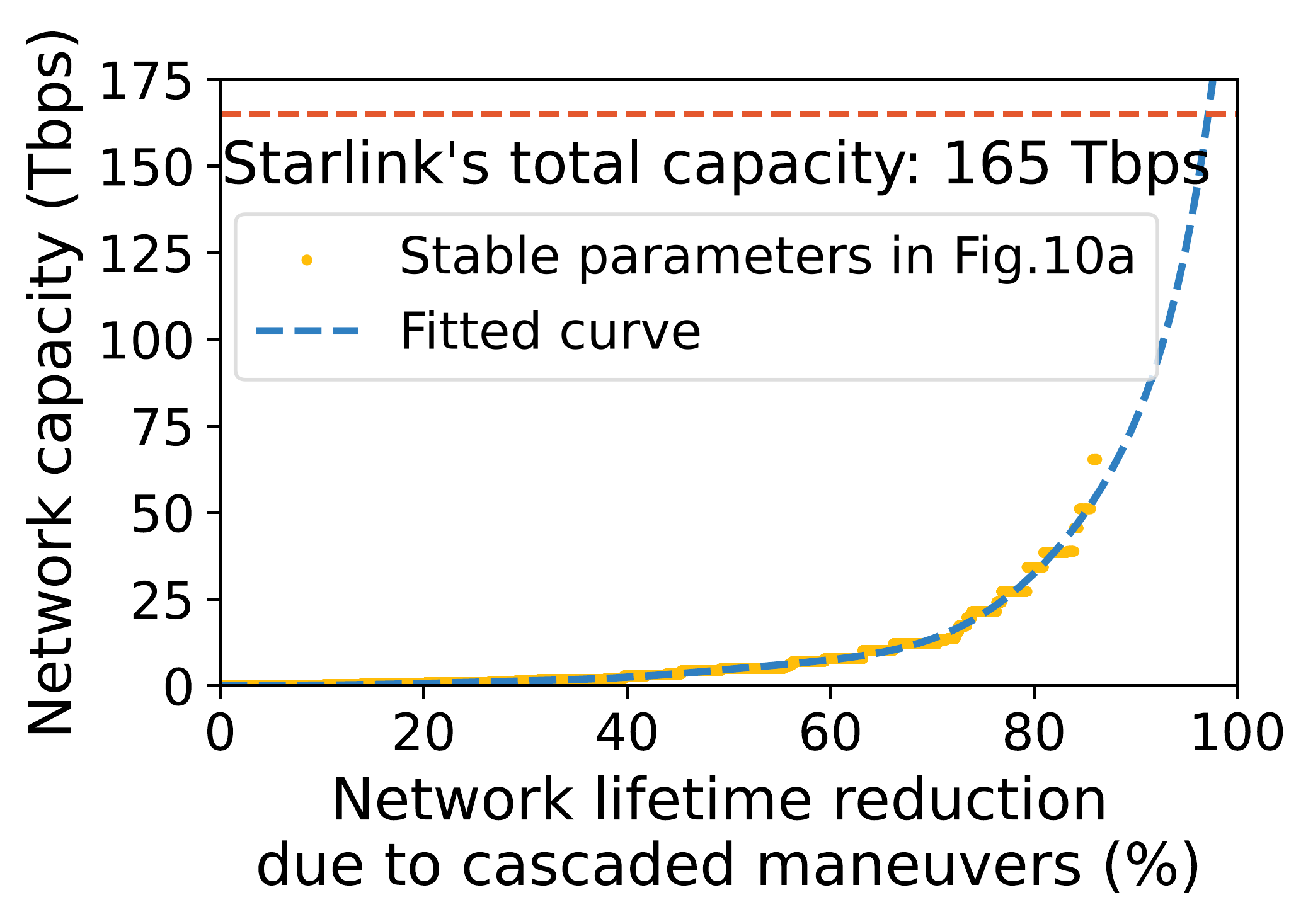}
	\label{fig:experimental-capacity-b}
	}
	\caption{Network capacity limit after stabilization.}
	\label{fig:experimental-capacity}
\end{figure}

Figure~\ref{fig:experimental-capacity} projects the LEO network capacity in different combinations of $(\alpha_1,\alpha_3)$ by fixing $\alpha_2$ as the most common value in Figure~\ref{fig:exp-policy-parameters}. 
It confirms the feasibility of fine-tuning Starlink's maneuver parameters for stabilization.
However, the maximal LEO network capacity after the stabilization would be highly constrained. 
This tradeoff between the LEO network capacity and lifetime is quantified in Figure~\ref{fig:experimental-capacity-b}.
It hints that Starlink may prioritize the short-term network capacity over the long-term network lifetime. 
This strategy may help accommodate more customers but also increase satellite replacement costs to offset Starlink’s net income. 

\section{Mitigation: Bilateral Control}
\label{sec:solution}

In this section, we explore how to renovate the self-driving mega-constellations to {\em simultaneously} boost LEO network lifetime and capacity. 
As shown in $\S$\ref{sec:theory}--\ref{sec:validation}, this goal is hard to achieve in the state-of-the-art local pairwise maneuver paradigm due to its tight coupling of two different tasks:

\begin{compactenum}

\item {\em Global stability:}
The inter-satellite motion transfer and amplification should be sufficiently small to avoid cascaded maneuvers for a longer network lifetime
($\S$\ref{sec:theory:network-lifetime});

\item {\em Local collision avoidance:} 
The inter-satellite spacing should be sufficiently large to 
tolerate each satellite's catchup to its preceding maneuvering satellite,
which can limit the LEO network capacity
($\S$\ref{sec:theory:network-capacity}).

\end{compactenum}
The local pairwise maneuver uses the same control law in Equation~\ref{eqn:collision-transfer}--\ref{eqn:collision-transfer-function} for both tasks, 
thus suffering from their conflicting satisfication conditions between LEO network lifetime (preferring $|H(\lambda)|<1$) and capacity ($|H(\lambda)|\geq1$).


Instead, these two tasks can be independent, decoupled, and governed by different control laws.
Although stability requires a small {\em forward} transfer function in Equation~\ref{eqn:collision-transfer}--\ref{eqn:collision-transfer-function} from the leading to the following satellite, 
collision avoidance does not mandate so.
As an alternative, collision avoidance can also be achieved {\em backward} by accelerating the leading satellite to prevent its successor's catchup.
This backward control is orthogonal and additive to the forward control in Equation~\ref{eqn:collision-transfer}--\ref{eqn:collision-transfer-function}, thus offering a chance of bypassing the LEO network lifetime-capacity dilemma for concurrent boosts.


To this end, we decouple LEO network lifetime extension from network capacity expansion via bilateral maneuver control, as shown in Figure~\ref{fig:bilateral-system}.
This method is inspired by the recent bilateral brake control in terrestrial autonomous driving \cite{horn2013suppressing, horn2017wave} but customized to be satellite-specific.
Unlike the legacy local pairwise policy in Figure~\ref{fig:theoretical-model},
each satellite's internal collision avoidance decision depends on its relative distance and velocity to not only its predecessor but also its successor.
This method incrementally adds a backward control to the existing forward control in local pairwise maneuvers.
It uses the combined backward/forward control for collision avoidance while still retaining stability using the forward control.
We next detail its control-theoretic model to prove its benefits for networking ($\S$\ref{sec:solution:model}),
discuss its practical deployment ($\S$\ref{sec:solution:deployment}),
and assess its viability in Starlink ($\S$\ref{sec:solution:evaluation}).

\subsection{Control-Theoretic Model}
\label{subsec:defense-bilateral}
\label{sec:solution:model}

\begin{figure}[t]
	\centering
	\vspace{-1mm}
	\includegraphics[width=\columnwidth]{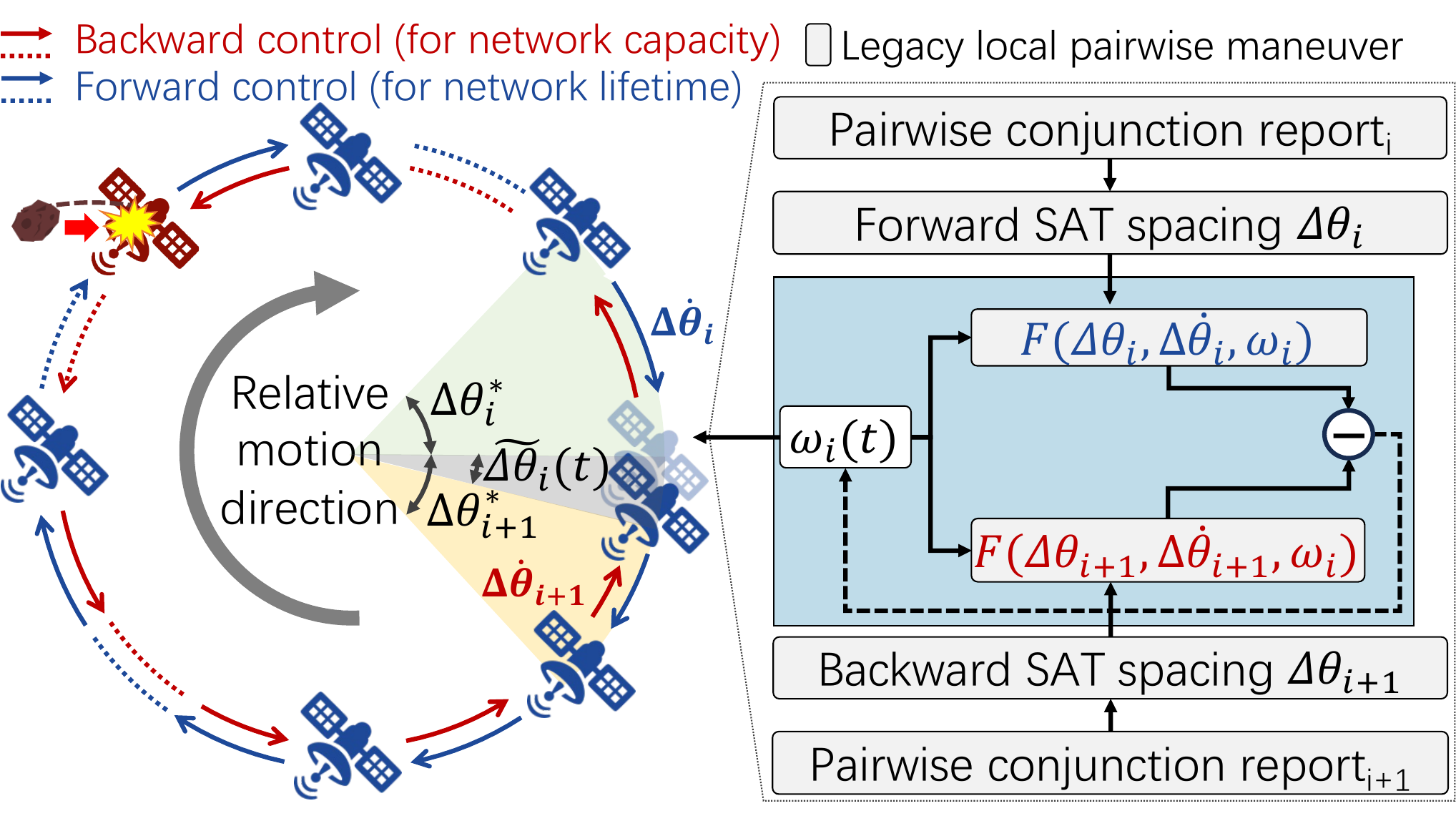}
	\caption{Bilateral maneuver control (video at \cite{bilateral-control-annimation}).}
	\label{fig:bilateral-system}
	\vspace{-2mm}
\end{figure}

Figure~\ref{fig:bilateral-system} overviews our bilateral maneuver control in the self-driving mega-constellation networks.
It follows the same modeling of cascaded collision avoidance in $\S$\ref{sec:theory:model}-\ref{sec:theory:lyapunov-analysis}, except that the $i$-th satellite's internal collision avoidance policy is renovated as
\begin{eqnarray}
\dot{\omega}_i(t)&=&F(\Delta\theta_i(t),\dot{\Delta\theta}_i(t),\omega_i(t))\nonumber\\
&-& F(\Delta\theta_{i+1}(t),\dot{\Delta\theta}_{i+1}(t),\omega_i(t))
\label{eqn:bilateral-control-policy}
\end{eqnarray}
Clearly, this bilateral control policy reuses and combines the existing local pairwise maneuver policy in Equation~\ref{eqn:self-driving-system} in a backward plus forward manner.
By replacing Equation~\ref{eqn:self-driving-system} with this new policy and repeating the analytics procedures in $\S$\ref{sec:theory:lyapunov-analysis}--\ref{sec:theory:network-lifetime}, we get its bilateral transfer function as 
\begin{equation}
\widetilde{\omega}_{i}(t)=\hat{H}(\lambda)\widetilde{\omega}_{i-1}(t)+\hat{H}(\lambda)\widetilde{\omega}_{i+1}(t)
\label{eqn:bilateral-control-transfer}
\end{equation}
where
\begin{equation}
\hat{H}(\lambda)=\frac{\alpha_1+\alpha_3\lambda}{2\alpha_1+2\alpha_3\lambda+\lambda^2}
\label{eqn:bilateral-control-transfer-function}
\end{equation}
is the collision transfer function in this bilateral control, and $\widetilde{\omega}_{i}(t)$, $\alpha_1$, and $\alpha_3$ are identical to those in $\S$\ref{sec:theory:lyapunov-analysis}--\ref{sec:theory:network-lifetime}.
Unlike pairwise maneuvers in Equation~\ref{eqn:collision-transfer}--\ref{eqn:collision-transfer-function},
a maneuvering satellite's motion $\widetilde{\omega}_{i}(t)$ in bilateral control depends on not only its leading satellite's motion $\widetilde{\omega}_{i-1}(t)$ but also its following satellite's motion $\widetilde{\omega}_{i+1}(t)$.
Intuitively, this offers an additional flexibility to avoid collisions {\em and} stabilize maneuvers\footnote{An informal physical analogy from \cite{horn2013suppressing, horn2017wave} to this bilateral control is a chain of satellites in Figure~\ref{fig:bilateral-system} inter-connected by ``springs and dampers.''
Any satellite's deviation from the equilibrium for collision avoidance will propagate both forward and backward, thus being attenuated and dissipated by the hop-by-hop dampers over time. 
This contrasts with amplifications in local pairwise maneuvers in Figure~\ref{fig:theoretical-model}, which have no such physical analogy.
}:

\begin{compactenum}
\item {\em Global stability:} To avoid cascaded maneuvers, the $i$-th satellite's motion should not be amplified when propagated to its following $(i+1)$-th satellite, \ie, $|\hat{H}(\lambda)|<1$.

\item {\em Local collision avoidance:} 
To avoid the $(i+1)$-th satellite from catching up with the maneuvering $i$-th satellite to cause internal collisions, the $i$-th satellite's overall gain in Equation~\ref{eqn:bilateral-control-transfer} should be large enough as $2|\hat{H}(\lambda)|\geq1$.

\end{compactenum}
Obviously, 
both conditions
can be satisfied simultaneously if $0.5\leq|\hat{H}(\lambda)|<1$, thus bypassing the LEO network lifetime-capacity dilemma in $\S$\ref{sec:theory:network-capacity} as follows:
\begin{prop}[Bilateral maneuver control]
\label{prop:stability-condition-bilateral-control}
For any $\alpha_1>0,\alpha_3>0$,
the bilateral control in Equation~\ref{eqn:bilateral-control-policy} 
always ensures 
the following properties simultaneously in the satellite ring
\begin{compactitem}
\item {\bf Network lifetime:}
The maneuvers in Equation~\ref{eqn:bilateral-control-policy} are always stable, thus suppressing cascaded maneuvers for longer satellite network lifetime; and
\item {\bf Network capacity:} Any positive inter-satellite spacing $\Delta\theta_{safe}>0$ suffices to avoid internal collisions, posing no constraints on the satellite network capacity.
\end{compactitem}
\end{prop}

Proposition~\ref{prop:stability-condition-bilateral-control} is proved in Appendix~\ref{proof:stability-condition-bilateral-control}.
Different from Proposition~\ref{prop:stability-condition} and Equation~\ref{eqn:safety-distance} in the local pairwise maneuver paradigm,
Proposition~\ref{prop:stability-condition-bilateral-control} 
ensures stability for all maneuver policies
{\em without} limiting the mega-constellation scale. 
This concurrently boots LEO network lifetime and capacity.


\subsection{Practical Deployment}
\label{sec:solution:deployment}

\begin{figure}[t]
\vspace{-14mm}
\centering
\subfloat[
Local pairwise maneuver
]{
\includegraphics[width=0.50\columnwidth]{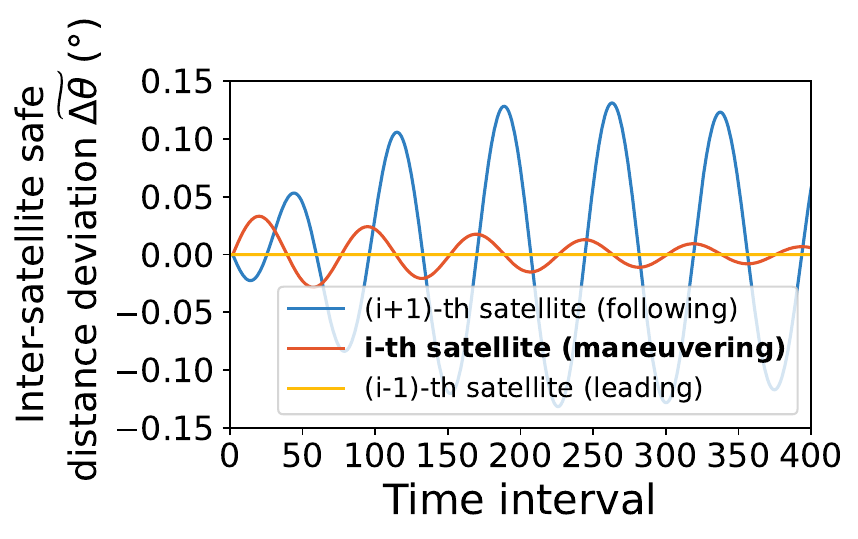}
\label{fig:pairwise-theta}
}
\subfloat[
Bilateral maneuver control]{
\includegraphics[width=0.5\columnwidth]{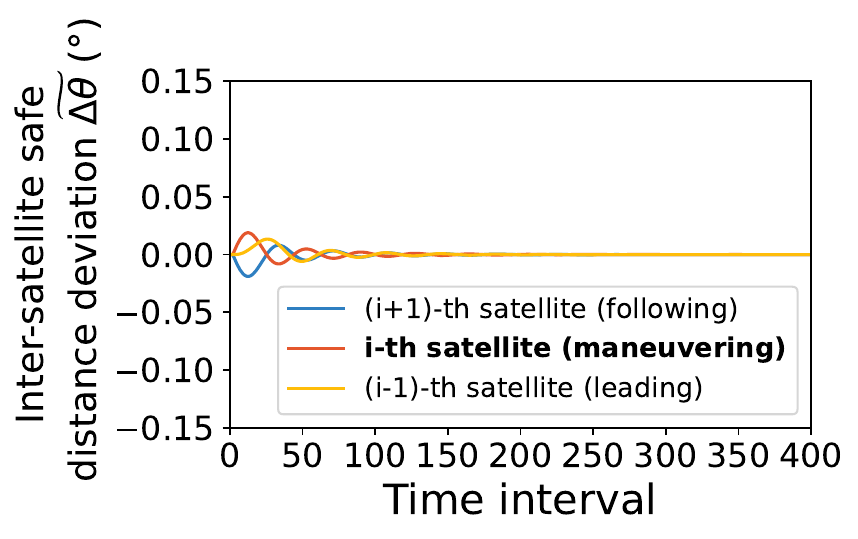}
\label{fig:bilateral-theta}
}
\vspace{-1mm}
\caption{
A showcase of different maneuvers' stability.
}
\label{fig:solution-bilateral}
\vspace{-2mm}
\end{figure}

Our bilateral control in Figure~\ref{fig:bilateral-system} is naturally backward compatible and incrementally deployable to existing self-driving mega-constellations.
Note that bilateral control is a simple linear combination of two legacy local pairwise maneuver policies. 
To adopt it, each satellite can directly reuse its legacy maneuver system in Figure~\ref{fig:starlink-ssa} to dry-run its legacy pairwise maneuver decision separately for its leading and following neighboring satellite inside the mega-constellation. 
Then, it follows Equation~\ref{eqn:bilateral-control-policy} to combine both dry-runs' outputs as its final maneuver decision.
No modifications of the legacy pairwise maneuver policy in Equation~\ref{eqn:self-driving-system}, satellite's hardware, standard pairwise conjunction reports, or ephemeris formats in Figure~\ref{fig:starlink-ssa} are needed for incremental deployment. 

Moreover, bilateral control can be readily realized as a distributed solution.
Similar to local pairwise maneuvers in Figure~\ref{fig:starlink-ssa}, each satellite only needs its neighboring satellites' information for local bilateral control without centralized scheduling, thus scalable to self-driving mega-constellations.

\subsection{Operational Trace-Driven Evaluation}
\label{sec:solution:evaluation}

\begin{figure}[t]
\vspace{-4mm}
\centering
\subfloat[
Local pairwise maneuver
]{
\includegraphics[width=0.50\columnwidth]{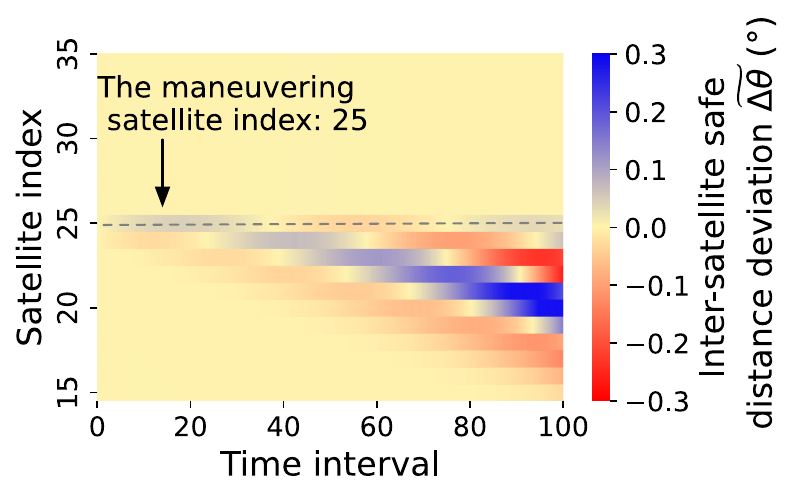}
\label{fig:pairwise-overtime}
}
\subfloat[
Bilateral maneuver control]{
\includegraphics[width=0.5\columnwidth]{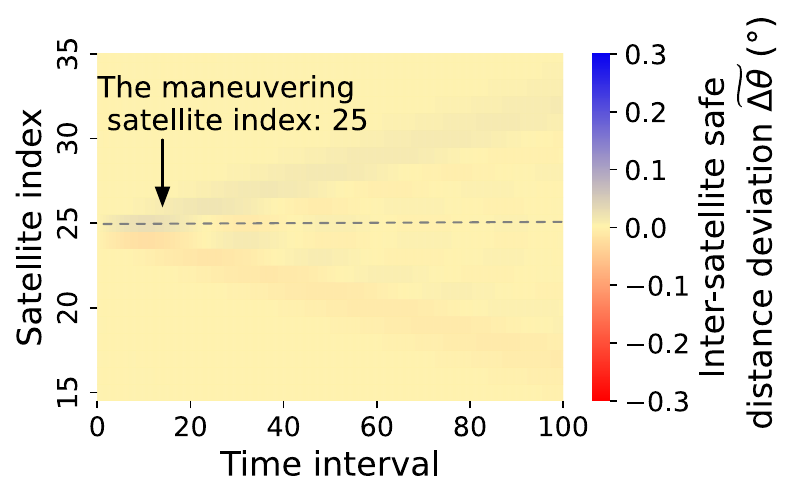}
\label{fig:bilateral-overtime}
}
\vspace{-1mm}
\caption{
Satellite spacing limiting network capacity.
}
\label{fig:solution-bilateral-overtime}
\vspace{-2mm}
\end{figure}

We assess the viability of bilateral control using operational trace-driven emulations.
Since bilateral control is a linear combination of local pairwise maneuvers, we emulate it by reusing Starlink's maneuvers in $\S$\ref{sec:validation:theory} and composing them via Equation~\ref{eqn:bilateral-control-policy}.
This allows for a fair comparison between bilateral control and existing maneuvers in the same setting. 

We first analyze how bilateral control mitigates cascaded collision avoidance to extend the network lifetime.
Figure~\ref{fig:solution-bilateral} showcases three neighboring satellites' deviation from the ideal constellation layout over time when the middle satellite (marked red) maneuvers for external collision avoidance.
If using Starlink's unstable local pairwise maneuvers in $\S$\ref{sec:validation:theory}, 
this maneuvering satellite will reduce its inter-satellite safe distance to its following neighbor only, which in turn has to amplify maneuvers for its own safety.
Instead, bilateral control distributes the maneuvering satellite's inter-satellite spacing changes across its leading and following neighbors.
This way, both neighboring satellites only need minor maneuvers for safety, thus suppressing hop-by-hop amplifications to save maneuver budgets for a longer network lifetime. 

We next assess bilateral control's impacts on the LEO network capacity.
Figure~\ref{fig:solution-bilateral-overtime} compares the temporal evolution of inter-satellite spacing in the holistic mega-constellation using different maneuver policies.
It confirms that bilateral control retains stable inter-satellite spacing by spreading and dissipating an external collision avoidance maneuver's significant safe distance changes hop by hop, as exemplified in Figure~\ref{fig:solution-bilateral}.
This implies that a small inter-satellite spacing is sufficient for bilateral control to avoid internal collisions inside the mega-constellation, thus removing the legacy maneuver's LEO network capacity bottleneck in Equation~\ref{eqn:leo-capacity}.

\begin{figure}[t]
\vspace{-13mm}
\centering
\subfloat[
Maneuver amplification factor
]{
\includegraphics[width=0.45\columnwidth]{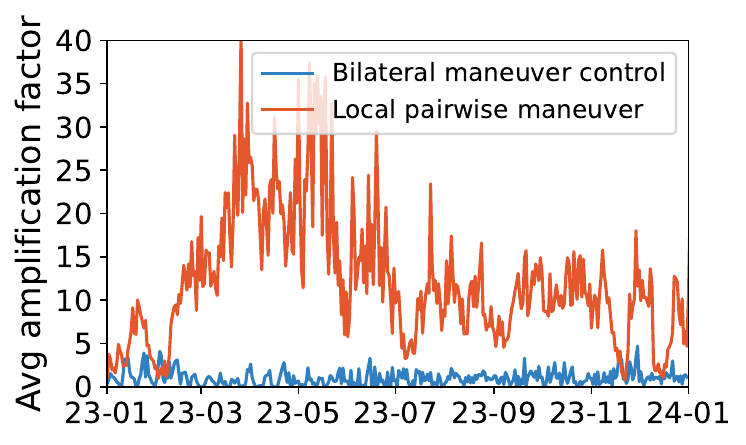}
\hspace{-2mm}
\includegraphics[width=0.195\columnwidth]{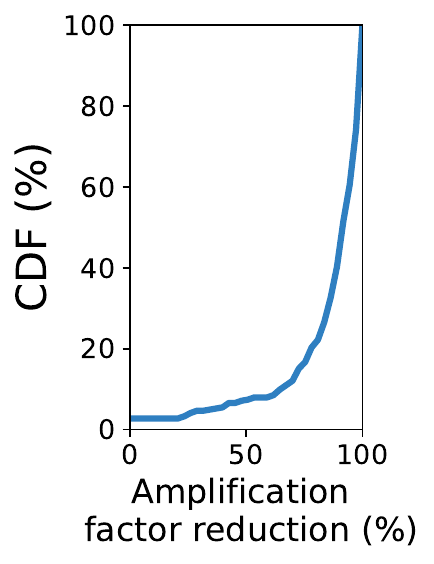}
\label{fig:solution-af}
}
\subfloat[
Lifetime extension]{
\includegraphics[width=0.315\columnwidth]{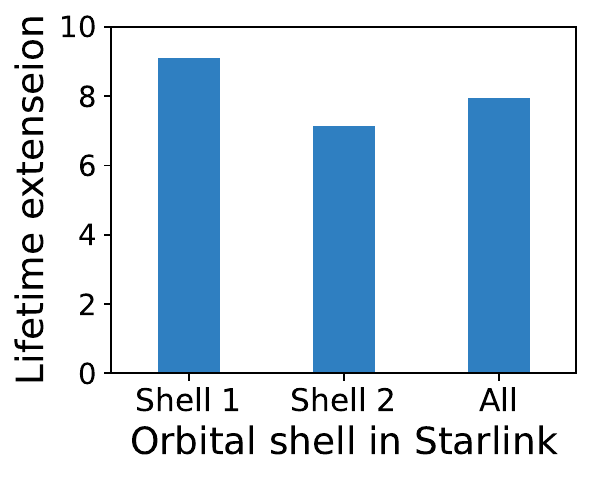}
\label{fig:solution-lifetime}
}
\caption{
Network lifetime in different maneuvers.
}
\label{fig:solution-compare}
\vspace{-2mm}
\end{figure}

With both merits, bilateral control can simultaneously benefit the LEO network's lifetime and capacity.
Figure~\ref{fig:solution-af} replays Starlink's external collision risk events from 2022/12 to 2023/12 and counts cascaded maneuvers induced by the legacy local pairwise maneuver in $\S$\ref{sec:validation:theory} and bilateral control atop it.
In Starlink's orbital shell 1 (at the 550 km altitude with 53\textdegree\ orbit inclination) and 2 (540 km altitude and 53.2\textdegree\ inclination),
bilateral control reduces local pairwise maneuvers' amplification factor by \add{80.41\%} on average, 
leading to an \add{8}$\times$ network lifetime extension without reducing the capacity.

\section{Related Work}
\label{sec:related}

Recent mega-constellation deployment has spurred academic and industrial enthusiasm for satellite networks
in diverse aspects,
including radio optimization \cite{jia2021uplink, jia2022analytic, singhspectrumize, pan2023pmsat}, 
topology \cite{bhattacherjee2019network,li2021cyber},
routing \cite{mobicom24li,uyeda2022loon,barritt2018loon,handley2019using,handley2018delay,lai2023achieving},
function split \cite{li2022spacecore,nsdi24liu},
scheduling \cite{vasisht2021l2d2,tao2023transmitting},
transport control \cite{hu2023leo,cao2023satcp,barbosa2023comparative},
orbital computing \cite{kassas2020navigation,bhattacherjee2020orbit, denby2020orbital,mobicom24xing},
positioning \cite{narayana2020hummingbird, iannucci2022fused,neinavaie2021acquisition,neinavaie2021exploiting}, 
and security \cite{smailes2023watch, willbold2023space, koisser2024don, sp24liu}.
Instead, the underlying orbital maneuvers 
gained less attention as 
they were outside the traditional networking research scope.
It is not until recently that \cite{mobicom23li,leonet23zhao} noticed maneuvers' impacts on topology.
We take one step forward to reveal their inherent instability and impacts on network lifetime and capacity. 

LEO mega-constellations are well known to congest orbits and raise collision risks \cite{kessler2010kessler}, which motivated self-driving systems for collision avoidance \cite{starlink-maneuver, starlink-maneuver-2,nasa-starling,probe2022prototype,brown2014simulated}.
Instead, we show that such systems themselves can threaten space safety. 
This phenomenon is analogous to ``phantom traffic jams'' in terrestrial driving \cite{sugiyama2008traffic,wu2018stabilizing,zheng2015stability} 
but more detrimental due to its additional impacts on network lifetime and capacity. 
Hence, we customize bilateral control \cite{horn2013suppressing, horn2017wave} for mega-constellations to strive for sustainable and performant satellite networking.






\section{Conclusion}
\label{sec:concl}

This paper studies how a self-driving mega-constellation's endogenous instability hinders its sustainable high-speed Internet services to numerous users.
We formally prove and empirically validate that, while initially designed for safety, 
its self-driving system itself can threaten
safety {and} networking with cascaded collision avoidance.
This domino effect is rooted in the decades-old {\em local pairwise} maneuver paradigm, which is reasonable for standalone satellites but unstable when scaled out to recent mega-constellation networks.
We devise bilateral maneuver control to mitigate it for concurrent satellite network lifetime {and} capacity boosts.
More refinements of network-friendly maneuvers can be explored in the future, 
such as enhancing its stability in more complex settings,
cross-layer optimizations for network functions,
and cooperative maneuvers between networked satellites.
We hope these lessons can help satellite operators
strive for a safer and faster
planet-scale high-speed Internet from space.

\medskip
\paragraphb{Acknowledgment:}
{
This work is funded by the National Key Research and Development Plan of China (2022YFB3105201) and the National Natural Science Foundation of China (62202261). Yuanjie Li and Hewu Li are the corresponding authors.}

\bibliographystyle{unsrt}
\bibliography{bib/collision,bib/extreme-mobility,bib/proposal,bib/satellite,bib/standard,bib/wing,bib/cellular,bib/security}

\appendix

\begin{appendix}

\section{Pairwise Space Collision Risk Assessment}
\label{appendix:pc}

\begin{figure}[t]
\includegraphics[width=\columnwidth]{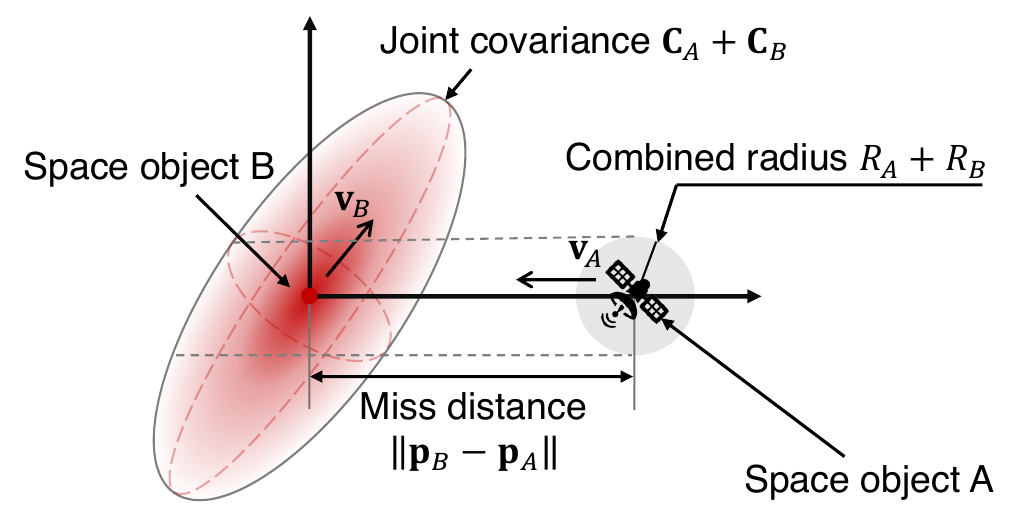}
\caption{The calculation of collision probability $P_c$ between two space objects on the conjunction plane.}
\label{fig:cdm}
\end{figure}

We follow \cite{krage2020nasa, operator-handbook, alfano2005numerical} to introduce the standard calculation of miss distance and collision probability between two space objects (satellites or debris). 
Consider a pair of independent space objects $A$ and $B$ that may collide, as shown in Figure~\ref{fig:cdm}.
To plan orbital maneuvers, satellite operators should estimate their miss distance and 
collision probability $P_c$ at the time of closest approach (TCA).
To compute it, we need to know both objects' size $R_k$, velocity $\mathbf{v}_k$, and position $\mathbf{p}_k\:(k=A,B)$. 
Note that, due to the inaccurate observations by radars/telescopes, both objects' velocity and position observations can be inaccurate.
This inaccuracy can cause random uncertainties about these objects' potential collision.  
The deviations of objects' velocity and position are 
normally distributed with a covariance matrix $\mathbf{C}_k\:(k=A,B)$.
Since the observation errors on the two objects are independent, 
we can shift all the errors into one imagined ``object'' of combined radius $R_A+R_B$ that passes through an ``error ellipsoid'' with a joint covariance matrix $\mathbf{C}_A+ \mathbf{C}_B$.
As visualized in Figure~\ref{fig:cdm}, this method reduces the 3D collision probability calculation into an equivalent 2D calculation on the {conjunction plane} perpendicular to the relative velocity vector. 
The collision probability  $P_c$  equals the probability that the miss-distance vector ``tube'' will cross this collision ``error ellipsoid''.

\begin{algorithm}[htp]
   \centering
        \begin{algorithmic}[1]
        \Require{Space object information $\{\mathbf{v}_k,\mathbf{C}_k,\mathbf{p}_k,R_k\}_{k\in\{A,B\}}$}
	\Ensure{Collision probability $P_c$ between space objects $A$ and $B$} 
	
	\State $\mathbf{v}_r\leftarrow\mathbf{v}_B-\mathbf{v}_A$ \textcolor{gray}{\Comment{{\em\tiny Relative velocity between $A$ and $B$}}}
	
	\State $\mathbf{\hat{e}_1}\leftarrow\frac{\mathbf{v}_r}{|\mathbf{v}_r|}$, $\mathbf{\hat{e}_2}\leftarrow\frac{\mathbf{v}_B\times\mathbf{v}_A}{|\mathbf{v}_B\times\mathbf{v}_A|}$, $\mathbf{\hat{e}_3}\leftarrow\mathbf{\hat{e}_1}\times\mathbf{\hat{e}_2}$, $\mathbf{Q}\leftarrow[\mathbf{\hat{e}_2} \quad \mathbf{\hat{e}_3}]$ \textcolor{gray}{\Comment{{\em\tiny Unit velocity vectors}}}
	
	\State $\mathbf{C}\leftarrow\mathbf{Q}^T(\mathbf{C}_A+\mathbf{C}_B)\mathbf{Q}$ \textcolor{gray}{\Comment{{\em\tiny Covariance (uncertainty) of space objects' trajectories}}}
	
	\State $(\mathbf{u,v})\leftarrow\text{Eigenvectors}(\mathbf{C})$, $(\mathbf{\sigma_x^2,\sigma_y^2})\leftarrow\text{Eigenvalues}(\mathbf{C})$ 
	
	\State $\mathbf{u}\leftarrow\frac{\mathbf{u}}{|\mathbf{u}|}$, $\mathbf{v}\leftarrow\frac{\mathbf{v}}{|\mathbf{v}|}$, 
	$\mathbf{U}\leftarrow[\mathbf{u} \quad \mathbf{v}]$
	
	\State $\begin{bmatrix}x_m\\y_m\end{bmatrix}\leftarrow\mathbf{U}^T\mathbf{Q}^T(\mathbf{p}_B-\mathbf{p}_A)$ \textcolor{gray}{\Comment{{\em\tiny Mean value of miss distance between $A$ and $B$}}}
	
	\State $R\leftarrow R_A+R_B$ \textcolor{gray}{\Comment{{\em\tiny Combined radius/size of space objects $A$ and $B$}}}

	\State Compute the pairwise collision probability $P_c$ between $A$ and $B$ as
	$$P_c=\frac{1}{2\pi\sigma_x\sigma_y}\iint_{x^2+y^2\leq R^2}f(x,y)dxdy$$
	where
	$$f(x,y)=\exp\left\{-\frac{1}{2}\left[\left(\frac{x-x_m}{\sigma_x}\right)^2+\left(\frac{y-y_m}{\sigma_y}\right)^2\right]\right\}$$
	
	\State \Return $P_c$
    \end{algorithmic}
    \caption{Collision probability between space objects.}
    \label{algo:collision-prob}
\end{algorithm}

Algorithm~\ref{algo:collision-prob} describes the standard collision probability calculation using the classic Alfano’s numerical method \cite{alfano2005numerical}. 
We first project the miss distance vector and the joint covariance to the 2D conjunction plane, as shown in Figure~\ref{fig:cdm}. 
The  $P_c$  is the probability that the miss-distance vector will cross the ``error ellipsoid.'' 
This is the portion of the combined error ellipsoid that falls within that circle. 
Therefore, the final collision probability is the integral of the Gaussian distribution probability density function in the circular domain (step 8). 

For our focus on internal collision avoidance inside each mega-constellation, 
both satellites' velocity and position are measured and gathered by the same satellite operator with high accuracy (\eg, using each Starlink satellite's onboard GPS in Figure~\ref{fig:starlink-ssa} \cite{starlink-maneuver, starlink-maneuver-2,narayana2020hummingbird}).
So, their uncertainties (\ie, covariance matrices $\mathbf{C}_A$ and $\mathbf{C}_B$ in Algorithm~\ref{algo:collision-prob}) are small or even negligible.
In this case, their collision probability calculation and the subsequent maneuver decision are dominated by these two satellites' miss distance (Line 6--7 in Algorithm~\ref{algo:collision-prob}).
To this end, our control-theoretic model in $\S$\ref{sec:motivation}--\ref{sec:solution} focuses on inter-satellite spacing $\Delta\theta$ inside the mega-constellation.


\section{Proof of Proposition~\ref{prop:stability-condition}}
\label{proof:stability-condition}

\begin{proof}

To asymptotically stabilize the linear self-driving system in Equation~\ref{eqn:linear-self-driving-system}, the Lyapunov stability theorem states that it should let all real parts of the complex-valued eigenvalues of the system matrix $A$ in Equation~\ref{eqn:linear-self-driving-system} be negative \cite{Lyapunov-stability}.
To compute the system matrix $A$'s eigenvalues, note that $A$ is a block circulant matrix, so it can be decomposed as follows according to \cite{marshall2004formations,olson2014circulant}:
$$
A=(F_n\otimes I_2)\cdot \text{diag}(D_1,D_2,...,D_n)\cdot(F_n^*\otimes I_2)
$$
where 
``$\otimes$'' denotes the Kronecker product on two matrices, 
``*'' denotes the conjugate transpose matrix,
$F_n$ is the $n\times n$ Fourier matrix defined as
$$
F_n=\frac{1}{\sqrt{n}}\begin{bmatrix}
1 & 1 & 1 & \cdots & 1\\
1 & z & z^2 & \cdots & z^{n-1}\\
1 & z^2 & z^4 & \cdots & z^{2(n-1)}\\
\vdots & \vdots & \vdots & \vdots & \vdots\\
1 & z^{(n-1)} & z^{2(n-1)} & \cdots & z^{(n-1)(n-1)}
\end{bmatrix}
$$
with $z=e^{\frac{2\pi j}{n}}$ and $j=\sqrt{-1}$, and 
$$
D_i=A_1+A_2z^{(n-1)(i-1)}=\begin{bmatrix}
0 & z^{(n-1)(i-1)}-1\\
\alpha_1 & \alpha_3z^{(n-1)(i-1)}-\alpha_2
\end{bmatrix}
$$
for $i=1,2,\cdots,n$. Then $A$'s eigenvalues can be calculated by 

\begin{eqnarray*}
\text{det}(\lambda I-A)&=&\text{det}(\lambda I-\text{diag}(D_1,D_2,...,D_n))\\
&=&\prod_{i=1}^n\text{det}(\lambda I-D_i)\\
&=&\prod_{i=1}^n\begin{vmatrix}
\lambda & 1-z^{(n-1)(i-1)}\\
-\alpha_1 & \lambda+\alpha_2-\alpha_3z^{(n-1)(i-1)}
\end{vmatrix}\\
&=&\prod_{i=1}^n\left[\lambda^2+\left(\alpha_2-\alpha_3z^{(n-1)(i-1)}\right)\lambda\right. \\
&&+\left.\alpha_1\left(1-z^{(n-1)(i-1)}\right)\right]\\
&=&0
\end{eqnarray*}
which implies that for $i=1,2,\cdots,n$
$$
\lambda^2+\left(\alpha_2-\alpha_3z^{(n-1)(i-1)}\right)\lambda+\alpha_1\left(1-z^{(n-1)(i-1)}\right)=0
$$
By substituting $z=e^{\frac{2\pi j}{n}}$ into this equation, we have
\begin{equation}
e^{\frac{i-1}{n}2\pi j}=\frac{\alpha_1+\alpha_3\lambda}{\alpha_1+\alpha_2\lambda+\lambda^2}=H(\lambda)
\label{eqn:lambda-root}
\end{equation}
where $H(\lambda)$ is exactly the collision transfer function derived in Equation~\ref{eqn:collision-transfer-function}.
This means that the eigenvalues of $A$ are also the solutions of Equation~\ref{eqn:lambda-root}.
Note that $e^{\frac{i-1}{n}2\pi j}$ is the $i$-th complex root of $z^n=1$.
Therefore, Equation~\ref{eqn:collision-transfer-function} indicates that for each eigenvalue $\lambda$ of $A$,
its value of $H(\lambda)$ corresponds to a unit root.
Therefore, if all the roots of $|H(\lambda)|=1$ have negative real parts, then the solutions of Equation~\ref{eqn:lambda-root} (\ie, all eigenvalues of $A$) have negative real parts.
This guarantees all eigenvalues of $A$ have negative real parts (\ie, $\text{Re}(\lambda)<0$) for the asymptotical stability according to the Lyapunov theory.
Note that this condition becomes sufficient and necessary for the case where the system is stable for any $n$, 
because now that $H(\lambda)$ can be any unit root $e^{\theta j}, \theta\in[0,2\pi)$ given all possible $n$ values.

To this end, we next derive the condition that all the roots of $|H(\lambda)|=1$ have negative real parts (\ie, $\text{Re}(\lambda)<0$).
To do so, we leverage the maximum modulus principle in complex analysis \cite{burckel2012introduction}.
Since the $\alpha_1$ and $\alpha_2$ are positive real numbers, the poles of $H(\lambda)$ are in the left half complex plane, indicating that $H(\lambda)$ is holomorphic in the right half complex plane.
Also, according to the definition of $H(\lambda)$ in Equation~\ref{eqn:collision-transfer-function}, $|H(\lambda)|\rightarrow0$ when $\text{Re}(\lambda)\rightarrow\infty$.
So according to the maximum modulus principle, 
the maximum of $|H(\lambda)|$ in the right half complex plane can only be achieved on the imaginary axis.
To avoid eigenvalues with positive real parts, $|H(\lambda)|$ should be no more than 1 on the imaginary axis, \ie,
$$
|H(j\mu)|\leq1, \forall \mu\in\mathbb{R}
$$
The necessary and sufficient condition to achieve this is
\begin{eqnarray*}
|H(j\mu)|^2=\frac{\alpha_1+j\alpha_3\mu}{\alpha_1+j\alpha_2\mu-\mu^2}\cdot\frac{\alpha_1-j\alpha_3\mu}{\alpha_1-j\alpha_2\mu-\mu^2}\leq1,\forall \mu\in\mathbb{R}
\end{eqnarray*}
Then we have
$$
\alpha_2^2-\alpha_3^2-2\alpha_1+\mu^2\geq0,\forall \mu\in\mathbb{R}
$$
Since $\mu^2\geq0,\forall \mu\in\mathbb{R}$, we conclude that to guarantee all eigenvalues of $A$ have negative real parts, it is necessary and sufficient to have
$$
\alpha_2^2-\alpha_3^2-2\alpha_1\geq0
$$
thus concluding our proof.
\end{proof}
\section{Proof of Proposition~\ref{prop:stability-condition-bilateral-control}}
\label{proof:stability-condition-bilateral-control}

\begin{proof}
We first derive the bilateral transfer function in Equation~\ref{eqn:bilateral-control-transfer}--\ref{eqn:bilateral-control-transfer-function}.
Consider a small perturbation of the $i$-th satellite from the equilibrium:
$$
\widetilde{\Delta\theta}_i(t)=\Delta\theta_i(t)-\Delta\theta^*,\:\:\:\:  \widetilde{\omega}_i(t)=\omega_i(t)-\omega^*
$$
Then we can apply the first-order Taylor expansion to linearize this satellite's collision avoidance policy in Equation~\ref{eqn:self-driving-system} at the equilibrium state $x_e=(\Delta\theta^*,\omega^*)$ as
\begin{eqnarray*}
\dot{\widetilde{\Delta\theta}}_i(t)&=&\omega_{i-1}(t)-\omega_i(t)= {\widetilde\omega}_i(t-1)-{\widetilde\omega}_i(t) \label{eqn:perturbation:theta-bilateral}\\			
\dot{\widetilde{\omega}}_i(t)&=&\alpha_1\left(\widetilde{\Delta\theta}_i(t)-\widetilde{\Delta\theta}_{i+1}(t)\right)\\
&+&\alpha_3\left(\left(\widetilde{\omega}_{i-1}(t)-\widetilde{\omega}_{i}(t)\right)-\left(\widetilde{\omega}_{i}(t)-\widetilde{\omega}_{i+1}(t)\right)\right)
\end{eqnarray*}
where $\alpha_1=\frac{\partial F}{\partial \Delta\theta}\vert_{x_e}$ and $\alpha_3=\frac{\partial F}{\partial \dot{\Delta\theta}}\vert_{x_e}$ are identical to those in $\S$\ref{sec:theory:lyapunov-analysis}.
To derive this linear system's collision transfer function, we repeat the procedures in $\S$\ref{sec:theory:network-lifetime} to consider the $i$-the satellite's motion deviation $\widetilde{\omega}_i(t)=c_ie^{\lambda t}$.
By substituting it into above Equations, we get the collision transfer function:
$$
\widetilde{\omega}_{i}(t)=\hat{H}(\lambda)\widetilde{\omega}_{i-1}(t)+\hat{H}(\lambda)\widetilde{\omega}_{i+1}(t)
$$
$$
\hat{H}(\lambda)=\frac{\alpha_1+\alpha_3\lambda}{2\alpha_1+2\alpha_3\lambda+\lambda^2}
$$
Next, consider a ring of $n$ satellites in Figure~\ref{fig:bilateral-system}.
Applying the above collision function to each hop yields:
\begin{eqnarray*}
\widetilde{\omega}_{1}(t)&=&\hat{H}(\lambda)\widetilde{\omega}_{n}(t)+\hat{H}(\lambda)\widetilde{\omega}_{2}(t)\\
\widetilde{\omega}_{2}(t)&=&\hat{H}(\lambda)\widetilde{\omega}_{1}(t)+\hat{H}(\lambda)\widetilde{\omega}_{3}(t)\\
&\cdots&\\
\widetilde{\omega}_{n-1}(t)&=&\hat{H}(\lambda)\widetilde{\omega}_{n-2}(t)+\hat{H}(\lambda)\widetilde{\omega}_{n}(t)\\
\widetilde{\omega}_{n}(t)&=&\hat{H}(\lambda)\widetilde{\omega}_{n-1}(t)+\hat{H}(\lambda)\widetilde{\omega}_{1}(t)
\end{eqnarray*}
By summing them up, we get 
$
\sum_{i=1}^n\widetilde{\omega}_{i}(t)=2\hat{H}(\lambda)\cdot\sum_{i=1}^n\widetilde{\omega}_{i}(t)
$
and thus 
$$
\hat{H}(\lambda)\equiv1/2
$$
By substituting Equation~\ref{eqn:bilateral-control-transfer-function} into it, we get
$$
\lambda\equiv0,\:\:\:\forall \alpha_1>0,\alpha_3>0
$$
According to Equation~\ref{eqn:bilateral-control-transfer}--\ref{eqn:bilateral-control-transfer-function}, this implies  
$$\widetilde{\omega}_{i}(t)\equiv\frac{1}{2}(\widetilde{\omega}_{i-1}(t)+\widetilde{\omega}_{i+1}(t)),\forall i=1,2,\cdots,n$$
Suppose $\widetilde{\omega}_{i-1}(t)\geq \widetilde{\omega}_{i}(t)$, then this equation immediately impies that
$$
\widetilde{\omega}_{i-1}(t)\geq \widetilde{\omega}_{i}(t)=\frac{1}{2}(\widetilde{\omega}_{i-1}(t)+\widetilde{\omega}_{i+1}(t))
$$
which further implies $\widetilde{\omega}_{i-1}(t)\geq \widetilde{\omega}_{i+1}(t)$.
Then we have
$$
\widetilde{\omega}_{i}(t)=\frac{1}{2}(\widetilde{\omega}_{i-1}(t)+\widetilde{\omega}_{i+1}(t))\geq \frac{1}{2}(\widetilde{\omega}_{i+1}(t)+\widetilde{\omega}_{i+1}(t))=\widetilde{\omega}_{i+1}(t)
$$
So by recursion on each hop, we get
$$
\widetilde{\omega}_{1}(t)\geq \widetilde{\omega}_{2}(t)\cdots\geq \widetilde{\omega}_{n}(t)\geq \widetilde{\omega}_{1}(t)
$$
which implies 
\begin{equation}
\widetilde{\omega}_{1}(t)= \widetilde{\omega}_{2}(t)\cdots= \widetilde{\omega}_{n}(t)= \widetilde{\omega}_{1}(t),\:\:\:\dot{\omega}_i(t)=0,\:\:\:\ \forall i
\label{eqn:bilateral-control:lifetime}
\end{equation}
\ie, under any self-driving policy parameters $\alpha_1>0, \alpha_3>0$, all satellites' motions remain homogeneous despite any small perturbations by external collision avoidance. This implies the stability and elimination of cascaded maneuvers in $\S$\ref{sec:theory:network-lifetime}.

We next analyze the bilateral control's satellite network capacity.
By repeating the analysis procedures in $\S$\ref{sec:theory:network-capacity}, we get
\begin{equation}
\dot{\widetilde{\Delta\theta}}_i(t)={\widetilde\omega}_i(t-1)-{\widetilde\omega}_i(t)\equiv0
\label{eqn:bilateral-control:safe-distance-change}
\end{equation}
So the minimum safe distance required by the LEO satellite mega-constellation to avoid internal collisions is
\begin{equation}
\Delta\theta_{safe}>\max_i\Delta\theta_i = \max_i\int_0^{\infty}\dot{\widetilde{\Delta\theta}}_i(t)dt\equiv0
\label{eqn:bilateral-control:safe-distance}
\end{equation}
\ie, while being stable, the bilateral control's minimum safe distance $\Delta\theta_{safe}$ can still be arbitrarily small, thus posing no limits on the satellite network capacity.
This contrasts with the existing local pairwise policy in Equation~\ref{eqn:safety-distance}
and bypasses its fundamental dilemma between the satellite network lifetime and capacity, thus concluding our proof.
\end{proof}

\end{appendix}

\end{document}